\documentclass[preprint2,longabstract]{aastex}

\newcommand{\ChaI}{Chamaeleon~I}

\shorttitle{Disk and Stellar Mass}
\shortauthors{Pascucci et al.}

\begin{document}


\title{A Steeper than Linear Disk Mass-Stellar Mass Scaling Relation}


\author{I. Pascucci\altaffilmark{1}}
\affil{Lunar and Planetary Laboratory, The University of Arizona, Tucson, AZ 85721, USA}
\altaffiltext{1}{Earths in Other Solar Systems Team, NASA Nexus for Exoplanet System Science}
\email{pascucci@lpl.arizona.edu}

\author{L. Testi\altaffilmark{2,3}}
\affil{European Southern Observatory, Karl-Schwarzschild-Strasse 2, D-85748 Garching bei M\"unchen, Germany}
\altaffiltext{2}{INAF-Arcetri, Largo E. Fermi 5, I-50125 Firenze, Italy}
\altaffiltext{3}{Gothenburg Center for Advance Studies in Science and Technology, Chalmers University of Technology and University of Gothenburg, SE-412 96 Gothenburg, Sweden}

\author{G. J. Herczeg and F. Long}
\affil{Kavli Institute for Astronomy and Astrophysics, Peking University, Yi He Yuan Lu 5, Haidian Qu, 100871 Beijing, China}

\author{C. F. Manara}
\affil{Scientific Support Office, Directorate of Science, European Space Research and Technology Centre (ESA/ESTEC), Keplerlaan 1, 2201 AZ Noordwijk, The Netherlands}

\author{N. Hendler and G. D. Mulders\altaffilmark{1}}
\affil{Lunar and Planetary Laboratory, The University of Arizona, Tucson, AZ 85721, USA}

\author{S. Krijt\altaffilmark{1} and F. Ciesla\altaffilmark{1}}
\affil{Department of the Geophysical Sciences, The University of Chicago, Chicago, IL 60637, USA}

\author{Th. Henning}
\affil{Max Planck Institute for Astronomy, K\"onigstuhl 17, D-69117 Heidelberg, Germany}

\author{S. Mohanty and E. Drabek-Maunder}
\affil{Imperial College London, 1010 Blackett Lab, Prince Consort Rd., London SW7 2AZ, UK}

\author{D. Apai\altaffilmark{1,4}}
\affil{Steward Observatory, The University of Arizona, Tucson, AZ 85721, USA}
\altaffiltext{4}{Lunar and Planetary Laboratory, The University of Arizona, Tucson, AZ 85721, USA}

\author{L. Sz\H{u}cs}
\affil{Max-Planck-Institut f\"ur extraterrestrische Physik, Giessenbachstrasse 1, D-85748 Garching, Germany}

\author{G. Sacco}
\affil{INAF-Arcetri, Largo E. Fermi 5, I-50125 Firenze, Italy}

\author{J. Olofsson}
\affil{Instituto de Fisica y Astronomia, Facultad de Ciencias, Universidad de Valparaiso, Playa Ancha, Valparaiso, Chile}

\begin{abstract}
The disk mass is among the most important input parameter for every planet formation model to determine the number and masses of the planets that can form.
We present an ALMA 887\,\micron{} survey of the disk population around objects from $\sim 2$ to 0.03\,$M_\sun$ in 
the nearby $\sim$2\,Myr-old \ChaI{} star-forming region.
We detect thermal dust emission from 66 out of 93 disks, spatially resolve 34 of them, and identify two disks with large dust cavities of
about 45\,AU in radius. 
Assuming isothermal and optically thin emission, we convert the 887\,\micron{} flux densities into dust disk masses, hereafter $M_{\rm dust}$.  We find that the $M_{\rm dust}-M_*$ relation is steeper than linear and of the form $M_{\rm dust} \propto (M_*)^{1.3-1.9}$, where the range in the power law index reflects two extremes of the possible relation between the average dust temperature and stellar luminosity. By re-analyzing all millimeter data available for nearby regions in a self-consistent way, we show that the 1-3\,Myr-old regions of Taurus, Lupus, and \ChaI{} share the same $M_{\rm dust}-M_*$ relation,  while the 10\,Myr-old Upper~Sco association has a steeper relation. Theoretical models of grain growth, drift, and fragmentation reproduce this trend and suggest that disks are in the fragmentation-limited regime. In this regime millimeter grains will be located closer in around lower-mass stars, a prediction that can be tested with deeper and higher spatial resolution ALMA observations. 
\end{abstract}





\section{Introduction}
The number of known exoplanets has exponentially grown in the past decade, revealing systems that are unlike our Solar System (e.g. \citealt{wf15}). While there is clearly a large diversity in planetary architectures, several trends with the mass of the central star are emerging. These include: i) a positive correlation between stellar mass and the occurrence rate of Jovian planets within a few AU (e.g. \citealt{johnson10,howard12,bonfils13}), although no correlation is present for the population of hot Jupiters within a 10\,days period \citep{obermeier16} and ii) a larger occurrence rate of close-in Earth-sized planets around M dwarfs than around sun-like stars \citep{dc13,mulders15a}. 
These trends are likely the result of stellar mass-dependent disk properties.
Indeed, planet formation models find that the disk mass strongly impacts the frequency and location of planets that can form, from giants down to Earth-size (e.g. \citealt{raymond07,alibert11,mordasini12}). Therefore, the scaling of disk mass versus stellar mass will yield a stellar mass dependence for the planet population.

Measuring gas disk masses is notoriously challenging both in the early ($\sim 1-10$\,Myr) protoplanetary phase (e.g. \citealt{kamp11,miotello14}) and in the late debris disk phase (e.g. \citealt{pascucci06,moor15}).
The disk mass in solids, up to mm-cm in size, is better constrained via continuum  mm-cm wavelength observations since the  emission from most dust grains is optically thin at these wavelengths. Still, individual dust disk masses can have an order of magnitude uncertainty because the absolute value of the dust opacity, which depends both on the grain composition and size distribution, is not known (e.g. \citealt{beckwith00}). 

Pre-ALMA millimeter surveys of nearby star-forming regions provided dust disk masses for over a hundred young stars, primarily with K and early M spectral types (see \citealt{wc11} and \citealt{testi14} for reviews). In spite of a large scatter
in disk masses at any stellar mass, the data were consistent with a linear disk mass  ($M_{\rm dust}$) $-$ stellar mass ($M_*$) scaling relation \citep{andrews13,mohanty13}, as hinted  earlier on by the detection of a few bright disks around sub-stellar objects \citep{klein03,scholz06,harvey12}. However, these studies were dominated by upper limits below the M0 spectral type, meaning that they only probed the upper envelope of disk masses in the low stellar mass end. This left open the possibility of a steeper $M_{\rm dust}-M_*$ relation buried in the non-detections. This suspicion was corroborated by the observation that stellar accretion rates ($\dot{M}$), tracing the gas disk component, display a steeper dependence with stellar mass when the population of low-mass stars is well sampled (e.g. \citealt{natta06,fang09,rigliaco11,alcala14}). If the steeper relation is due to the way disks viscously evolve and disperse (e.g. \citealt{hartmann06,aa06,ercolano14}) and if $M_{\rm dust}$ somehow traces the total (gas+dust) disk mass,  $\dot{M}$ and $M_{\rm dust}$ should  scale similarly with stellar mass.

The increased sensitivity of ALMA is now enabling us to survey entire star-forming regions and to probe the millimeter luminosity of young ($\sim 1-10$\,Myr) protoplanetary disks identified in previous infrared images.  The 1.3\,mm survey of the Orion OMC1 detected continuum emission toward 49 cluster members and reported no correlation between $M_{\rm dust}$ and $M_*$ \citep{eisner16}. However, as also pointed out by the authors, the statistical significance of this result is limited given the small number of ALMA detections and that spectroscopically-determined stellar masses in the OMC1 are only available for less than half of the ALMA-detected sources. The survey of the $5-10$\,Myr old Upper Sco association \citep{slesnick08} covered  all known disks around stars from $\sim$0.15 to 1.5\,$M_\sun$ and reported a steeper than linear relation between $M_{\rm dust}$ and $M_*$  \citep{bare16}. After removing debris/evolved transitional disks, they also found that the $M_{\rm dust}/M_{*}$ ratio in Upper~Sco is 
$\sim$4.5 times lower than that in Taurus, suggesting that significant evolution occurs in the outer disk between 1 and 10\,Myr. Finally, \citet{ansdell16} carried out a similarly sensitive ALMA survey in the much younger ($\sim$1-3\,Myr) Lupus star-forming clouds, covering sources in the I to IV regions, which most likely trace different stages of disk evolution. 
One of the main results of the \citet{ansdell16} survey is that the $M_{\rm dust} - M_*$ relation in Lupus is similar to that in Taurus and shallower than that in Upper Sco.

Here, we present an ALMA 887\,\micron{} survey of the $\sim2$\,Myr-old \ChaI{} star-forming region targeting disks around objects ranging from 2\,$M_\sun$ down to the sub-stellar regime (Sections~\ref{sect:sample} and \ref{sect:obs}). We demonstrate that the $M_{\rm dust} - M_*$ relation in \ChaI{} is steeper than linear, under a broad range of assumptions made to convert flux densities into dust disk masses (Sections~\ref{sect:results} and \ref{sect:DustDisk}). By re-analyzing in a self-consistent way all the sub-mm fluxes and stellar properties available for other nearby star-forming regions we also show that Taurus, Lupus, and \ChaI{} have the same $M_{\rm dust} - M_*$ relation, within the inferred uncertainties, and confirm that the one in Upper~Sco is steeper (Section~\ref{sect:discussion}). We discuss the possibility that the steeper relation traces either the growth of pebbles into larger solids that become undetectable by ALMA or a more efficient inward drift in disks around the lowest mass stars (Section~\ref{sect:discussion}).

\section{The Chamaeleon~I sample}\label{sect:sample}
In previous studies our group has assembled the stellar properties and spectral energy distribution (SED) of each \ChaI{} member and used continuum radiative transfer codes to model disk structures down to the substellar regime \citep{sz10,mulders12,olofsson13}. Our modeling included optical, 2MASS, Spitzer, WISE, and, when available, Herschel and mm photometry.  We did not include any spectroscopic data, e.g. Spitzer IRS spectra. Only objects displaying excess emission
at more than one wavelength were included in our ALMA survey. In this way we excluded all Class~III objects \citep{luhmanetal08}. In addition, we removed the few known Class~0 and I sources \citep{luhmanetal08, belloche11}. These criteria result in 93 objects with dust disks, mostly Class~II, but see later for sub-groups. Table~\ref{tab:SourceProperties} includes their 2MASS designations, other commonly used names, multiplicity information from the literature, and the spectral types (SpTy) from \citet{luhman07, luhman08}. This latter information was also used to set the exposure times (see Section~\ref{sect:obs}). We note that our sample is not complete in the sub-stellar regime (SpTy later than M6). For instance, the well known disk around the M7.75 brown dwarf Cha~H$\alpha$1 (e.g. \citealt{pascucci09}) is not included in our ALMA survey. Our ALMA sample also includes 32 known multiple stars. Assuming an average distance of 160\,pc to the \ChaI{} star-forming region \citep{luhman08}, 7/32 are "close" binaries, with projected separations $\leq$40\,AU that are small enough to affect disk evolution \citep{kraus12}.


The SEDs of 87 of our ALMA targets are classified in \citet{luhmanetal08} and \citet{manoj11} using the spectral slope $\alpha = d {\rm log} (\lambda F_\lambda)/d {\rm log} \lambda$ between $\sim$2\,\micron{} (2MASS K-band photometry) and 24\,\micron{} ({\it Spitzer}/MIPS photometry in the first contribution and {\it Spitzer}/IRS spectroscopy in the second). As discussed in \citet{manoj11} the two SED classifications are in good agreement. Six of our ALMA targets\footnote{The six unclassified targets are: J11160287-7624533, J11085367-7521359, J10561638-7630530, J11071181-7625501, J11175211-7629392, and J11004022-7619280.} were not observed with {\it Spitzer} but all have WISE photometry at 12\,\micron{} (W3 channel, \citealt{cutri12}). We use the following approach to classify them. First, we plot the de-reddened\footnote{To de-redden the magnitudes we used the A$_{\rm J}$ extinctions provided in \citet{luhman07} and the \citet{mathis90} reddening law because all of our sources have low extinction, A$_{\rm J}<0.8$.} $\alpha_{2-24}$ versus $\alpha_{2-12}$ for all \ChaI{} members that have 2MASS K-band, WISE 12\,\micron, and MIPS 24\,\micron{} photometry. From this plot we find that the two quantities are well correlated and the best fit relationship is:
 $\alpha_{2-24} = 1.14 (\pm 0.03) \times \alpha_{2-12} + 0.38 (\pm 0.06)$. Hence, we use this relationship to compute $\alpha_{2-24}$ from the measured $\alpha_{2-12}$ for the 6 unclassified sources. The inferred $\alpha_{2-24}$ spectral indices are between -1.7 and -0.9, all Class~II SED following  \citet{manoj11}. 
 The transitional disks (Class~II/T) are identified as having a deficit of flux at wavelengths less than 8\,\micron{} compared with the Class~II median and comparable or higher excess emission beyond  $\sim$13\,\micron{} following \citet{kim09} and \citet{manoj11}. By excluding the IRS Spitzer spectroscopy from our analysis we missed the Class~II/T disk around the M0 star Sz~18, also known as T25 \citep{kim09}. Its infrared excess is only pronounced beyond $\sim 15$\,\micron{} and the source was outside the MIPS 24\,\micron{} field of view \citep{luhmanetal08}, thus appearing as a Class~III source based on the available photometry.
 In summary, our sample includes 3 flat spectra (FS), 82 Class~II, and 8 Class~II/T.
    
As part of a parallel effort to simultaneously derive stellar parameters, extinction, and mass accretion rates, our group has obtained VLT X-Shooter spectra for 89 out of 93 of our ALMA \ChaI{} targets. The observations, data reduction, and properties inferred from the VLT spectra are summarized in \citet{manara14,manara16a,manara16c}. For eight sources, typically late M dwarfs, these new spectra were either not acquired or lacked enough signal-to-noise (hereafter, S/N) to reliably derive stellar and accretion properties, hence we adopt here the spectral type classification and stellar properties reported in \citet{luhman07, luhman08}, see Table~\ref{tab:SourceProperties}. As discussed in \citet{manara16a,manara16c} the difference between the new and literature spectral type is in most cases less than a spectral subclass. The largest difference occurs for the K-type stars and is thought to arise from the lack of good temperature diagnostics in the low-resolution red spectra used in previous studies for spectral classification. 

\begin{figure}[h]
\centering
\includegraphics[width=0.5\textwidth]{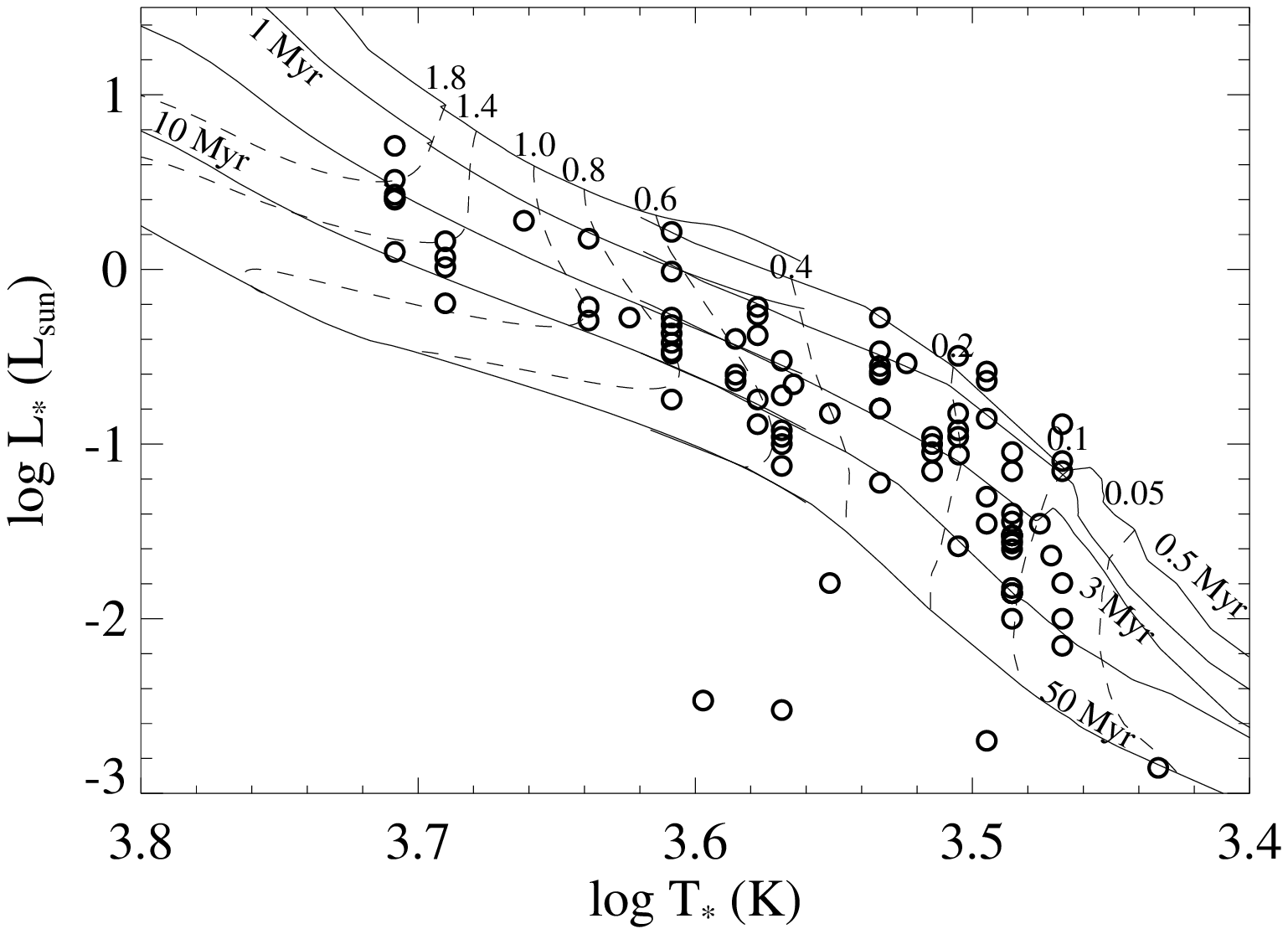}
\caption{H-R diagram of our ALMA \ChaI{} sample (each source is represented by an empty circle). The non-magnetic evolutionary tracks from \citet{feiden16} are plotted for effective temperatures greater than 3,700\,K (SpTy M1 and earlier) and masses greater than 0.5\,M$_\sun$. For effective temperatures lower than 4,200\,K and masses lower than 0.5\,M$_\sun$ we plot the evolutionary tracks from \citet{baraffe15}. Note the similarity of the two sets of isochrones in the overlapping effective temperature region for ages $\ge$1\,Myr.}
\label{fig:HRdiagram}
\end{figure}

\subsection{Stellar mass estimates} \label{sect:stellar_masses}
To derive stellar masses (and ages) we followed the standard approach of comparing empirical effective temperatures and stellar luminosities to those predicted by pre-main sequence evolutionary models. 
Effective temperatures and luminosities for our ALMA \ChaI{} sample are taken from \citet{manara14,manara16a,manara16c} and \citet{luhman07, luhman08} as summarized in column 9 of Table~\ref{tab:SourceProperties}.
The H-R diagram is shown in Figure~\ref{fig:HRdiagram} with each object represented by an empty circle and the evolutionary tracks from \citet{baraffe15} and the non-magnetic tracks from \citet{feiden16} in solid (isochrones) and dashed (stellar mass tracks) lines.

Our choice of evolutionary tracks is motivated by the recent work of \citet{hh15}  who demonstrated that these new models better match empirical stellar loci for low-mass stars and brown dwarfs in nearby young associations than older models. In addition, they yield very similar ages for low-mass stars (see Figure~4 in \citealt{hh15}), hence they can be combined to extend the stellar mass coverage. This is critical for our \ChaI{} sample which, as shown in Figure~\ref{fig:HRdiagram}, spans a large range in stellar mass, from above 1.5\,M$_\sun$ down to the substellar regime\footnote{The \citet{feiden16} tracks cover from 0.09 to 5.7\,M$_\sun$ while the \citet{baraffe15} from 0.015 to 1.4\,M$_\sun$, hence they are the only ones available in the sub-stellar regime.}. 

Following \citet{andrews13}, we adopt a Bayesian inference approach to assign a stellar mass, an age, and associated uncertainties to each of our ALMA targets. The first step in this approach is to interpolate the \citet{baraffe15} and \citet{feiden16} models on a common, finely sampled, age grid. Based on the \ChaI{} H-R diagram in Figure~\ref{fig:HRdiagram}, we include the earliest isochrones at 0.5\,Myr through to 50\,Myr-old isochrones with a step of 0.01 in log scale. Stellar masses are also sampled with the same spacing in log scale.  We use the \citet{baraffe15} tracks for all objects with effective temperatures $\le$3,900\,K (M dwarfs) and  switch to the \citet{feiden16} tracks for hotter stars (spectral types K and earlier). This procedure is motivated by the fact that around $\sim$3,900\,K the two sets of isochrones nicely overlap even for 1\,Myr-old stars (see Figure~4 in \citealt{hh15}), although there remains a small mismatch at the earliest 0.5\,Myr isochrone (see Figure~\ref{fig:HRdiagram}).

\begin{figure}[h]
\centering
\includegraphics[width=0.5\textwidth]{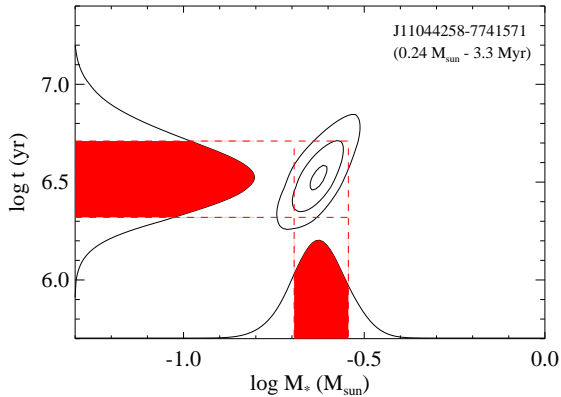}
\caption{Example of the likelihood function used to estimate stellar masses and ages. The best fit parameters for J11044258-7741571
are listed on the top right of the panel. The 68\% confidence intervals are the red regions of the marginal probability density functions.
These regions are calculated from the cumulative integral such that the area above and below the best fit parameter are each 0.34.}
\label{fig:exMassAge}
\end{figure}

For each ALMA target, identified by a temperature $T_{*}$ and luminosity $L_{*}$ in the H-R diagram, we compute a conditional likelihood function, assuming uniform priors on the model parameters,  as:
{\small
\begin{equation}\label{eq:likelihood}
F(\hat{T},\hat{L} \vert T_{*},L_{*}) = \frac{1}{ 2 \pi \sigma_{T_{*}} \sigma_{L_{*}} } exp (  -0.5 \times [ \frac{(T_{*} - \hat{T})^2 }{ \sigma_{T_{*}}^2 }  + \frac{(L_{*} -\hat{L})^2}{\sigma_{L_{*}}^2}  ]  )
\end{equation}
}
where $\hat{T}$ and $\hat{L}$ are the model grid temperatures and luminosities, while $\sigma_{T_{*}}$ and $\sigma_{L_{*}}$ are the uncertainties associated with $T_{*}$ and $L_{*}$. The uncertainty in log($T_{*}$) is assumed to be 0.02 for SpTy earlier than M3 and 0.01 for later SpTy while the uncertainty in log($L_{*}$) is taken to be 0.1 (see \citealt{manara16c}). We then integrate $F(\hat{T},\hat{L} \vert T_{*},L_{*})$ over the age and mass covered by the model grids and obtain two marginal probability density functions, see the curves in Figure~\ref{fig:exMassAge}. The best fit mass and age are the peaks of these functions and the uncertainties are the values that encompass 68\% of the area under the functions. 

This approach could be applied to all but 9 sources for which age estimates are found to be at the boundary of our grid. For the four sources for which our method identifies the youngest 0.5\,Myr isochrone and appear over-luminous in the H-R diagram\footnote{J11065906-7718535, J11094260-7725578, J11105597-7645325, and J11183572-7935548}, we choose this isochrone and compute the stellar mass based solely on the stellar effective temperature. For the other five sources\footnote{J10533978-7712338, J11063945-7736052, J11082570-7716396, J11111083-7641574, and J11160287-7624533} for which our method gives the oldest isochrone of 50\,Myr we take the median age of our \ChaI{} sources and again compute stellar masses based solely on stellar effective temperatures. Three  out of these five 'old' sources (J10533978-7712338,  J11111083-7641574, and  J11160287-7624533) have SED and/or spatially resolved imagery suggesting that the central star is surrounded by an edge-on disk  \citep{luhman07,robberto12}, thus explaining why they appear under-luminous in the H-R diagram.
We note that our ALMA sample has a median age of 3.5\,Myr, slightly older than the previously computed median age \citep{luhman07}. The resulting masses and their uncertainties, when available, are reported in the last column of Table~\ref{tab:SourceProperties}. 

\section{Observations and data reduction}\label{sect:obs}
Our observations were carried out as part of the ALMA Cycle~2 campaign on 2014 May 1-3 UTC (54 sources) and on 2015 May 18-19 UTC (39 sources). The 2014 observations included all stars with SpTy from Luhman equal or earlier than M3 (hereafter, {\it Hot} sample) while in 2015 we observed the remaining later SpTy sources (hereafter, {\it Cool} sample). 

All observations were obtained in Band 7 with a spatial resolution of $0.7''\times0.5''$, see Table~\ref{tab:ALMAObservations} for details on the number of 12m antennas, baselines, and calibrators. Each science block (SB), comprising either all {\it Hot} or {\it Cool} sources plus any calibrator, was executed twice.
The correlator was configured to record dual polarization with three continuum basebands of 5.6\,GHz aggregated bandwidth centered at 330.0, 341.1, and 343.0\,GHz for an average frequency of 338\,GHz (887\,\micron). The fourth baseband was devoted to the serendipitous detection of gas lines and was split in two sub-bands of 0.1\,GHz each centered at 329.3\,GHz and  330.6\,GHz to cover the C$^{18}$O (3-2) and the $^{13}$CO (3-2) transitions. This paper focuses on the continuum data, the reduction and analysis of the CO data will be presented in a separate contribution (Long et al. in prep.). Exposure times for the {\it Hot} sample were set to achieve a 1$\sigma$ rms of 1\,mJy/beam in the aggregated continuum bandwidth, while for the {\it Cool} sample we required 0.2\,mJy/beam. As a comparison previous single dish mm observations of the \ChaI{} star-forming region had 1$\sigma$ sensitivities greater than 10\,mJy over a beam of $\sim 20''$ \citep{henning93,belloche11}.

The ALMA data were calibrated using the CASA software package. The initial reduction scripts were provided by the North American ALMA Science Center and included phase, bandpass, and flux calibration. We re-ran the scripts using CASA  4.3.1. We used Pallas as the flux calibrator for the {\it Hot} sample SBs,  Ganymede for the first {\it Cool} sample SB, and the quasar J1107-448 for the second {\it Cool} sample SB. The flux scale was within 5\% and 8\% of the two SBs for the {\it Hot} and {\it Cool} samples respectively. For both samples, we used the average of the two SB fluxes in the calibration script. In the analysis that follows we adopt a conservative 1$\sigma$ uncertainty of 10\%{} on the absolute flux scale.

Dirty continuum images were created from the calibrated visibilities using CASA
v4.4 and natural weighting and by averaging the three continuum basebands (see
Figures~16 to 21 in the electronic version of the paper). We computed the rms of each image in a region outside the expected target location and found a median of 0.99 and 0.23\,mJy/beam for the {\it Hot} and {\it Cool} samples, very close to the requested sensitivities. We also computed an initial flux density at the target location by integrating within the 3\,rms closed contour. This flux density, in combination with the image rms and visual inspection, was used to decide if a source is detected. With this approach we classified 45/54 {\it Hot} and 21/39 {\it Cool} targets as detected. 

We also identified ten bright {\it Hot} and two bright {\it Cool} sources with S/N ranging from 36 to 100 and rms larger than 2 times the median rms that would benefit from self-calibration. 
For these 12 sources\footnote{The 10 {\it Hot} sources that require self-calibration are J10581677-7717170, J10590699-7701404, J11022491-7733357, J11040909-7627193, J11074366-7739411, J11080297-7738425, J11081509-7733531, J11092379-7623207, J11094742-7726290, and J11100010-7634578 while the two {\it Cold} sources are J11004022-7619280 and J11062554-7633418.} we 
followed the steps suggested by the North American ALMA Science Center for the brightest of our targets,  J11100010-7634578.
From each of the 12 measurement sets we produced an image with Briggs robust weighting parameter of zero and cell size 0\farcs075.  First, we performed a shallow cleaning on each image, down to a threshold of about 5 times the median rms of the {\it Hot} or {\it Cold} sample, and saved the model in the measurement set header. We then calibrated the phases using the model data column, applied the new calibration to the measurement set, and produced a new image from the better-calibrated data. We repeated the cycle of cleaning and phase calibration a second time starting from the new image and by applying a deeper cleaning, down to about 3 times the median rms of the {\it Hot} or {\it Cold} sample. The image produced in this second cycle was cleaned a third time, with phases and amplitudes calibrated and applied to the original measurement set. With this approach we found that the final image rms always improved,  reaching the median value of $\sim$1\,mJy/beam for the {\it Hot} and $\sim$0.2\,mJy/beam for the {\it Cold} samples even for the brightest of our sources, J11100010-7634578, whose initial image rms was $\sim$24\,mJy/beam. The 12 phase and amplitude calibrated measurement sets are used in the following steps to compute the source parameters.

\section{Results}\label{sect:results}
To compute the flux densities and to determine whether the emission is spatially resolved we rely on the visibility data as, e.g. discussed in \citet{carpenter14}.
First, we fit all of our 66 detections with an elliptical Gaussian using the {\it uvmodelfit} task in CASA. This model has 6 free parameters: the integrated flux density; the offsets in right ascension and declination from the phase center;
the FWHM; the aspect ratio;  and the position angle. With the underlying assumption that the model describes well the data, we scale the uncertainties on the fitted parameters by the factor needed to produce a reduced $\chi^2$ of 1. If the ratio of the FWHM to its uncertainty is less than 2, which happens for 32 sources, we also fit the visibility data with a point source model which has only 3 free parameters: the integrated flux density and  the offsets in right ascension and declination from the phase center. For 25 out of 32 sources we find that the reduced $\chi^2$ of the point source model is less than that of the Gaussian model, hence we adopt the point source fits. Even for the 7 sources where the reduced $\chi^2$ of the Gaussian model is lower than that of a point source model, we adopt the point source fits because the difference in the models' reduced $\chi^2$ is much smaller than the uncertainty on their values, which is approximately $\sqrt{2/N}$ for the over 7,000 visibility points that are fitted.
Finally, for the 27 sources that are not detected we also fit a point source model keeping the offsets in right ascension and declination fixed to -0\farcs3 and 0\farcs0, respectively, the median values from the sources that are detected. 

\begin{figure*}[h]
\centering
\includegraphics[width=1\textwidth]{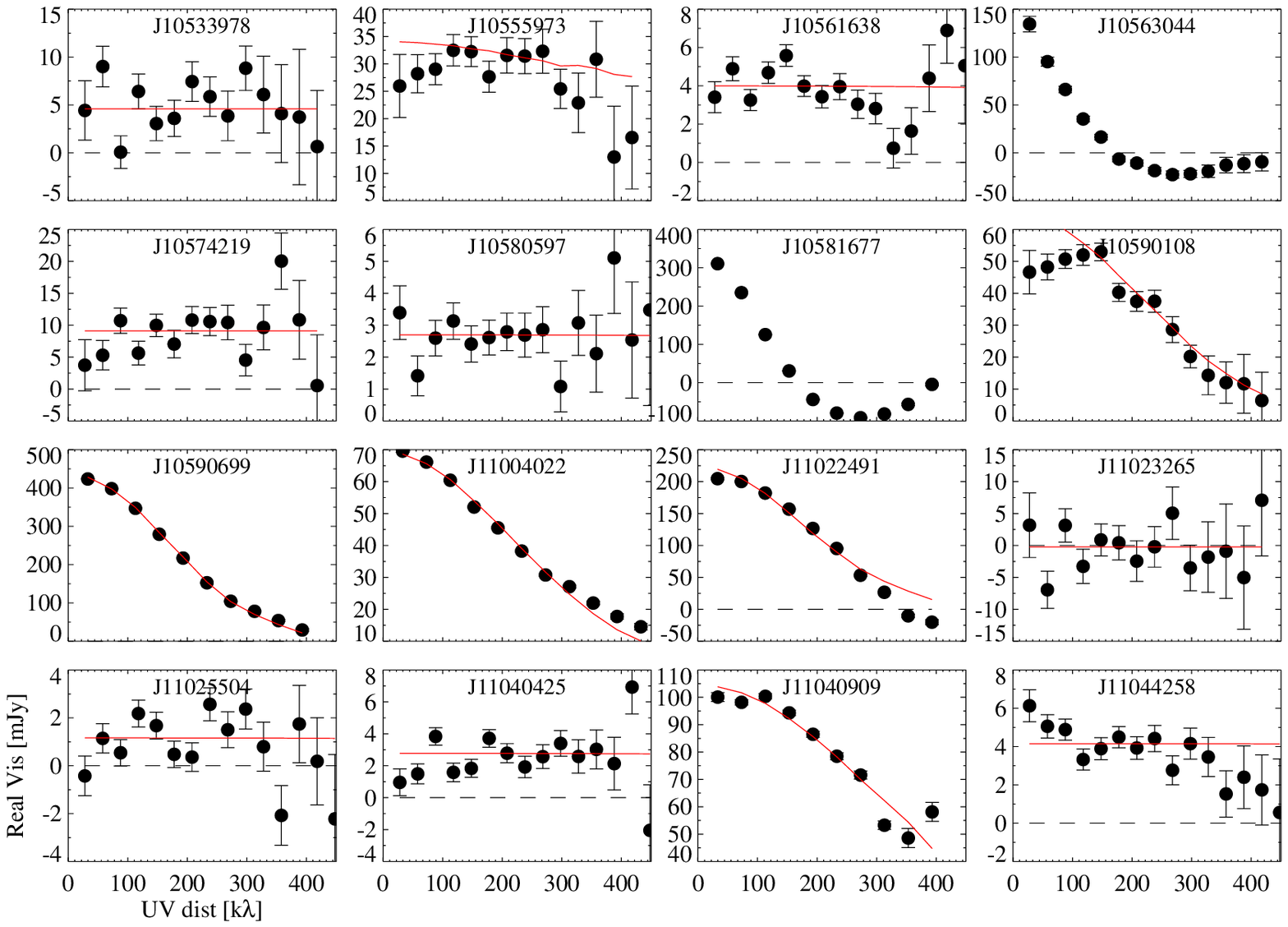}
\caption{Real part of the observed visibility (circles) as a function of the projected baseline using a sampling of 30\,k$\lambda$. The best model fit to the data (red solid line) is shown for all targets except the two disks with resolved cavities, see text for details. Similar figures for all other targets are available online.}
\label{fig:vis1}
\end{figure*}

\begin{figure*}[h]
\centering
 \begin{tabular}{ll}
 \includegraphics[angle=0,scale=0.45]{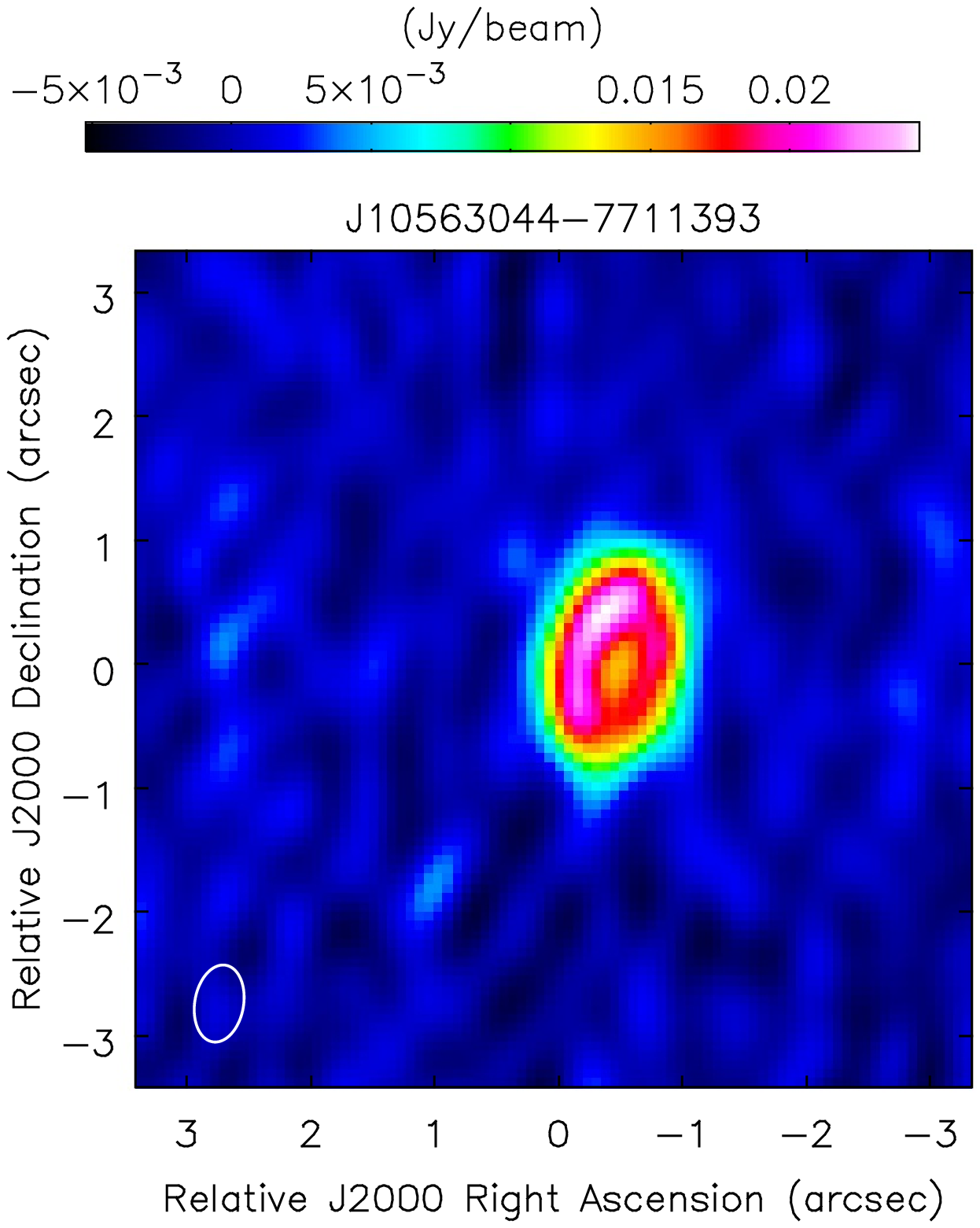} & 
 \includegraphics[angle=0,scale=0.45]{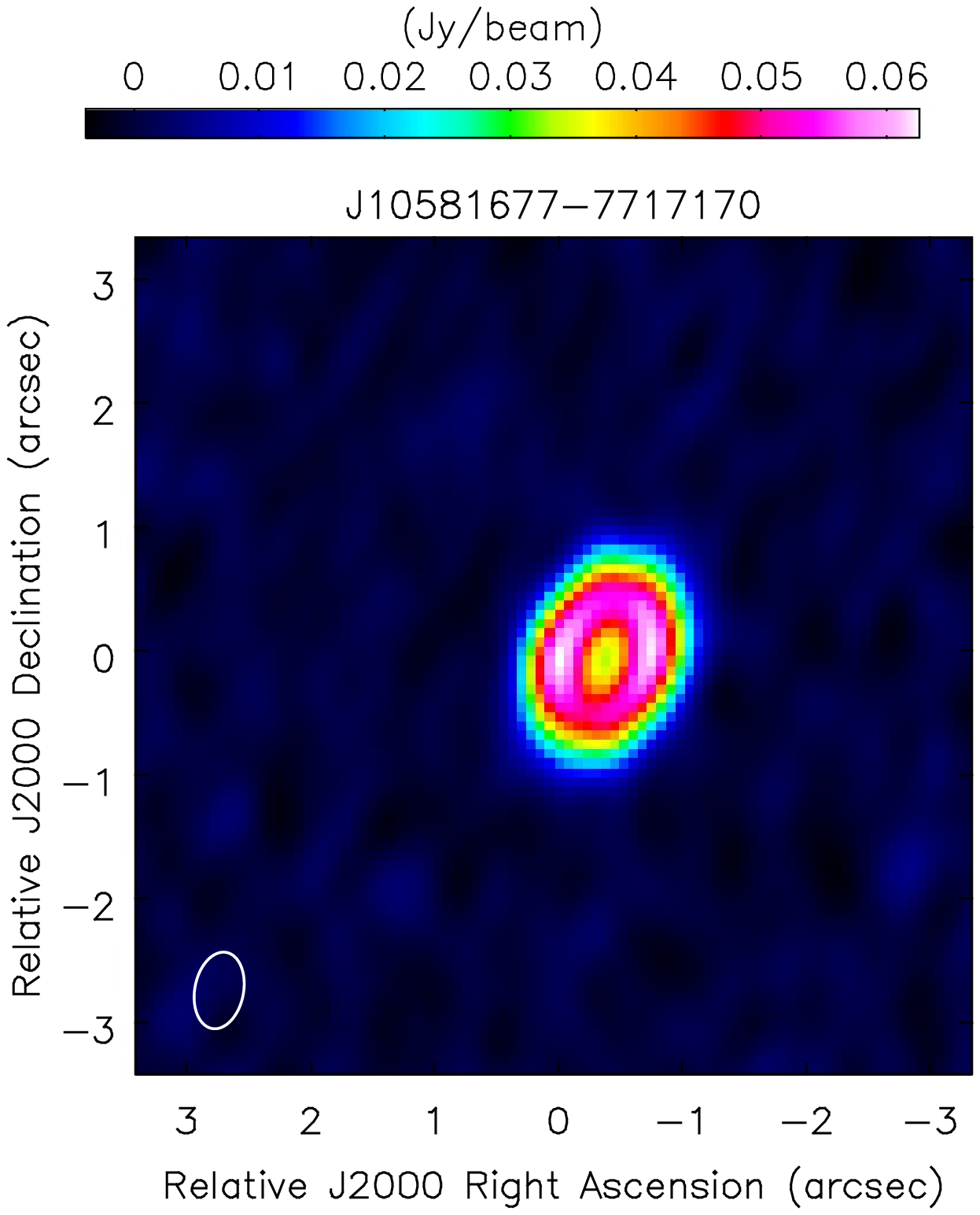} 
 \end{tabular}
 \vspace{-5 cm}
 \caption{The two disks in our \ChaI{} sample with spatially resolved dust cavities. J10581677 is a known transition disk based on its infrared photometry and spectroscopy.} \label{fig:cavity}
\end{figure*}

To visualize the goodness of the fits we compare the best fit model (solid line)
to the real component of the observed visibilities (filled circles) as a
function of projected baseline length (UV distance), see Figure~\ref{fig:vis1}
as an example, all other figures are available in the electronic version. In these figures all visibilities are re-centered to the continuum centroids found with {\it uvmodelfit}, each visibility point is the average of the visibilities within a 30\,k$\lambda$ range, and the error bars are the standard deviation divided by $\sqrt{N-1}$ where $N$ is the number of visibility points in the same range.
About half of the detected sources have spatially resolved emission, as evidenced by visibilities that decline in amplitude with increasing UV distance. 
Among them, J10563044-7711393  and J10581677-7717170 have resolved dust cavities, hence the Gaussian fit discussed above does not provide a good estimate for the source flux density. For these two sources we compute flux densities within the 3$\sigma$ contour in the deconvolve image\footnote{We remind the reader that J10581677-7717170 was one of the sources that required self-calibration (see Section~\ref{sect:obs}), hence the flux density is computed on the final phase and amplitude calibrated image.}, see Figure~\ref{fig:cavity}.  J10581677-7717170 is a known transition disk with an estimated dust cavity of $\sim$30\,AU in radius \citep{kim09}. On the contrary, J10563044-7711393 has not been classified as a transition disk based on its infrared photometry but a {\it Spitzer}/IRS spectrum could not be extracted for this source due to its faintness \citep{manoj11}.  The radius of both cavities is $\sim$45\,AU as measured from the images and from the location of the first null in the visibility plot (see eq. A9 in \citealt{hughes07}).

Overall, we have identified two sources with dust disk cavities, 32 sources whose mm emission is resolved (elliptical Gaussian model), 32 sources with unresolved mm emission (point source model), and 27 sources with too faint or absent mm emission to be detected in our survey. Among the resolved mm sources 23 belong to the {\it Hot} sample and 9 to the {\it Cool} sample implying that $\sim$51\% and 39\% of the detected sources are resolved in the two samples respectively. Table~\ref{tab:SourceFluxes} summarizes the measured continuum flux densities ($F_\nu$) and uncertainties, offsets from the phase center in right ascension and declination for the detected sources ($\Delta \alpha$ and $\Delta \delta$), and FWHMs for the resolved mm sources. In the analysis that follows we calculate upper limits for sources that are not detected as 3 times the uncertainty on $F_\nu$ which is also reported in Table~\ref{tab:SourceFluxes}.

\begin{figure}[h]
\centering
\includegraphics[width=0.5\textwidth]{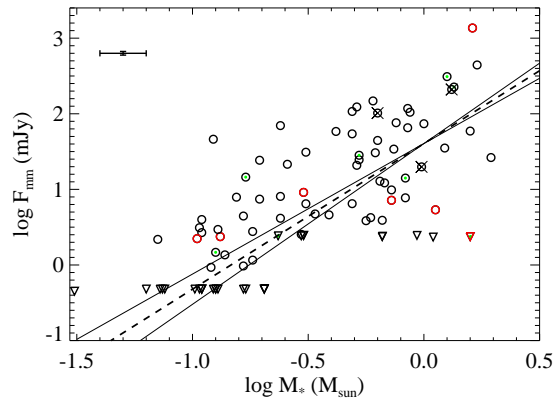}
\caption{Flux densities ($F_{\rm mm}$) as a function of stellar masses ($M_*$). Circles are sources with detected mm flux while downward pointing triangles represent non detections. Sources marked with an 'X' are FS disks, a green dot within the main symbol denotes Class II/T SEDs while a red color denotes 'close' binaries (projected separation $\leq$40\,AU). The dashed line gives the best fit relationship using a Bayesian approach that accounts for censored data. The median errorbar in log($M_*$) and log($F_{\rm mm}$) is shown in the upper left corner of the plot and corresponds to  $\pm 0.1$\,dex and $\pm 0.02$\,dex respectively.} 
\label{fig:FnuMstar}
\end{figure}

Flux densities and upper limits as a function of stellar masses are shown in Figure~\ref{fig:FnuMstar} in a log-log plot, circles for detections and downward pointing triangles for non-detections. 
Note that the SED-identified transition disks are not among the brightest mm disks. Two of them,
J11071330-7743498 (SpTy M3.5) and J11124268-7722230 (SpTy G8), remain undetected at our sensitivity. However, the latter source has also a  $\sim$0.7\,$M_\sun$ companion at a projected distance of 38\,AU \citep{daemgen13} that might have tidally truncated the disk of the primary, leading to a lower than average mm flux.  The disks around J11100704-7629376 and 
J11103801-7732399, two K-type stars with companions at $\sim 20$\,AU and 27\,AU distance respectively,
also appear fainter than disks around stars of similar stellar mass and might have been truncated. 
Stars in Taurus with companions at tens of AU have also fainter disks than expected for their mass \citep{harris12}.
At the other extreme, the star J11100010-7634578 has a companion at 65\,mas and the brightest mm disk, in this case a circumbinary disk. Circumbinary disks are also found to be among the brightest mm disks in Taurus \citep{harris12}.

Figure~\ref{fig:FnuMstar} demonstrates that mm fluxes have a spread of more than a dex at a given stellar mass, part of which, as mentioned above, may be attributed to stellar multiplicity. In spite of the spread, flux densities are strongly correlated with stellar mass. This trend is not unique to the \ChaI{} star-forming region \citep{andrews13,mohanty13,bare16,ansdell16}. Assuming a linear relationship in the log-log plane, we can determine the best fit using the Bayesian method developed by \citet{kelly07} that properly accounts for the measurement uncertainties, non-detections, and intrinsic scatter. This Bayesian method assumes Gaussian measurement errors, hence we have adopted the full range of the stellar mass uncertainty, covering 68\%{} of the area under the marginal probability density function, and divided it by two as the error on each stellar mass. For the 10 sources where we had to fix the isochrone we use the median uncertainty in log($M_*$) of $\pm0.1$\,dex. With this approach we find the following best fit relationship: log($F_{\rm mm}$/mJy)=1.9($\pm0.2$)$\times$log($M_*/M_\sun$)+1.6($\pm0.1$). 

Although our $1\sigma$ confidence intervals in stellar mass are often not symmetric around the best value, they are still small enough that the assumed Gaussian distribution does not affect the Bayesian fit. We tested that only when the error on log($M_*$) becomes larger than 2.5 times the median value, the best fit relation is no longer consistent with the one reported here and the inferred slope steepens. This means that the intrinsic scatter in mm fluxes drives the best fit given our measurement errors in log($F_{\rm mm}$) and log($M_*$).
This is also confirmed by other regression methods that do not account for measurement errors but recover the same slope and intercept of the Bayesian approach within the quoted uncertainties (see Appendix~\ref{sect:regression}).
To further test the robustness of this relation, we also compute stellar masses using the effective temperatures and luminosities in \citet{luhman07} and find the following best fit   log($F_{\rm mm}$/mJy)=2.1($\pm0.3$)$\times$log($M_*/M_\sun$)+1.7($\pm0.2$), basically consistent with the one using the new stellar properties. Thus, the $F_{\rm mm}-M_*$ relation in \ChaI{} is much steeper than linear and the mm flux scales almost with the square of the stellar mass.

The 1.9-2.1($\pm$0.2) slope of \ChaI{} is within the 1.5-2.0 range reported for Taurus, where the range in Taurus reflects the use of different evolutionary tracks to assign stellar masses \citep{andrews13}. For the old \citet{baraffe98} tracks, which in \citet{andrews13} are the most similar to the ones we use, the $F_{\rm mm} - M_{*}$ slope in Taurus is 1.5$\pm0.2$, lower but still marginally consistent with the one we find in \ChaI{}. We caution that lower values can also result from low sensitivity at the lower stellar mass end.  As a test we degrade our sensitivity to the typical 850\,\micron{} 1$\sigma$ sensitivity of  $\sim$3\,mJy achieved in Taurus, 3 and 15 times worse that the actual sensitivity of the {\it Hot} and {\it Cool} samples in \ChaI{}. The best fit slope of this degraded dataset is only 1.3$\pm$0.2, still consistent with the Taurus one but shallower than the slope we measure in \ChaI{} with the actual sensitivities. This simple test demonstrates the need for deep millimeter surveys to reveal the intrinsic disk flux$-$stellar mass dependence.

\section{Dust disk masses}\label{sect:DustDisk}
Dust disk emission at millimeter wavelengths is mostly optically thin, hence continuum flux densities can be used to estimate dust disk masses (e.g. \citealt{beck90}). We adopt the simplified approach commonly used in the field (e.g. \citealt{natta00}) and assume isothermal and optically thin emission to compute disk masses as follows:
{\small
\begin{equation}\label{eq:Mdisk}
log\,M_{\rm dust} = log\,F_\nu + 2\,log\,d - log\,\kappa_\nu -  log\,B_\nu(T_{\rm dust})
\end{equation}
}
where $F_\nu$ is the flux density at 338\,GHz (887\,\micron), $d$ is the distance (160\,pc for \ChaI, \citealt{luhman08}), $\kappa_\nu$ is the dust opacity, and  $B_\nu(T_{\rm dust})$ is the Planck function at the temperature $T_{\rm dust}$.
We adopt a dust opacity of 2.3 cm$^2$\,g$^{-1}$ at 230\,GHz with a frequency dependence of $\nu^{0.4}$, the same as in  \citet{andrews13} for Taurus and \citet{carpenter14} for Upper~Sco. 
The average dust temperature responsible for the mm emission ($T_{\rm dust}$) is poorly constrained.  \citet{andrews13} performed 2D continuum radiative transfer calculations for a representative grid of disk models and proposed the following scaling relation for stars in the 0.1 to 100\,$L_\sun$ luminosity range: $T_{\rm dust}=25$K$\times (L_*/L_\sun)^{0.25}$. However,  \citet{van16} and \citet{hendler16} show that a weaker $T_{\rm dust} - L_*$ dependence can be reached by adjusting some of the disk input parameters used in \citet{andrews13}, most notably the outer disk radius.  In particular, \citet{hendler16} find that  if lower mass stars have smaller dust disks then the $T_{\rm dust} - L_*$ relation flattens out, becoming almost independent of stellar luminosity if the dust disk radius scales linearly with stellar mass. As discussed in Section~\ref{sect:results},  the percentage of resolved disks is higher in the {\it Hot} than in the {\it Cool} sample, perhaps hinting on smaller dust disks around lower mass stars. However, this could be also due to low S/N at the low end of
the stellar mass spectrum.  Because a S/N on the continuum $\ge 30$ is needed to properly estimate dust disk sizes \citep{tazzari16}, deeper ALMA observations are needed to pin down if and how the disk size scales with stellar mass.
 
Given the uncertainty in the $T_{\rm dust} - L_*$ relation,  we compute dust disk masses for two extreme cases:  a) a constant $T_{\rm dust}$ fixed to 20\,K to directly compare our results to recent ALMA surveys of other star-forming regions (e.g. Lupus, \citealt{ansdell16})  and b) a varying $T_{\rm dust}$ with stellar luminosity as proposed by \citet{andrews13}.
{Several studies have applied a plateau of $\sim $10\,K to the outer disk temperature \citep{mohanty13,ricci14,testi16}, given that this is the value reached by dust grains heated by the interstellar radiation field in giant molecular clouds \citep{mathis1983}. We have decided not to apply this plateau in our study for two reasons. 
First, continuum radiative transfer models show that the interstellar radiation field has a negligible effect on the dust disk temperature and outer disks can be colder than 10\,K  \citep{van16,hendler16}. Second, \citet{guilloteau16} note that the edge-on disk of the Flying Saucer absorbs radiation from CO background clouds and infer very low dust temperatures of 5-7\,K at $\sim$100\,AU in this disk. The lowest luminosity source in our \ChaI{} sample, J11082570-7716396 with $L_{\rm bol}=0.0014\,L_\sun$, has a $T_{\rm dust}$ of 4.8\,K with our prescription. Such a value is below 10\,K but still consistent with the lower temperatures found in disk models and in the Flying Saucer disk.

Figure~\ref{fig:MdiskMstar} summarizes our findings with black and orange symbols for case a) and b) respectively.
A lower $T_{\rm dust}$ for lower luminosity (typically lower mass) objects results in a lower Planck function hence in a higher dust mass estimate. When applying to these two extreme cases the same Bayesian approach described in Section~\ref{sect:results} we find the following best fits: 
\\
log($M_{\rm dust}/M_\Earth$)=1.9($\pm0.2$)$\times$log($M_*/M_\sun$)+1.1($\pm0.1$) for a constant $T_{\rm dust}$ and 
\\
log($M_{\rm dust}/M_\Earth$)=1.3($\pm0.2$)$\times$log($M_*/M_\sun$)+1.1($\pm0.1$) for $T_{\rm dust}$ decreasing with stellar luminosity.  The standard deviation (hereafter, dispersion) about the regression is 0.8$\pm 0.1$\,dex, see also Table~\ref{tab:relations}. As expected, the slope of the $M_{\rm dust} - M_*$ relation is the same as that of the $F_{\rm mm}$--$M_*$ relation for the assumption of constant temperature while it is flatter when the temperature decreases with stellar luminosity.
Importantly, even the flatter relation is steeper than the linear one inferred from pre-ALMA disk surveys \citep{andrews13,mohanty13} and from ALMA surveys with a limited coverage of stellar masses (e.g. \citealt{carpenter14} and Section~\ref{sect:results}). Most likely the $M_{\rm dust} - M_*$ relation is steeper than 1.3($\pm0.2$) since our 
ALMA observations as well as recent analysis of brown dwarf disks hint at smaller dust disks around lower mass stars \citep{testi16,hendler16}. However, quantifying the steepness of the $M_{\rm dust} - M_*$ relation will require measuring how dust disk sizes scale with stellar masses.

\begin{figure}[h]
\centering
\includegraphics[width=0.5\textwidth]{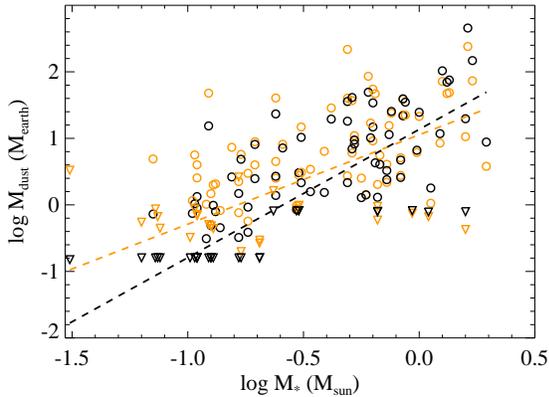}
\caption{Dust disk masses ($M_{\rm dust}$) as a function of stellar masses ($M_*$). Black symbols are for a constant dust disk temperature of 20\,K while orange symbols use the $T_{\rm dust} - L_*$ scaling relation proposed by \citet{andrews13}. The dashed lines are the best fits for these two cases. Note that the scaling relation proposed by \citet{andrews13} flattens the disk-stellar mass relation.
} 
\label{fig:MdiskMstar}
\end{figure}

\section{Discussion}\label{sect:discussion}

\subsection{The disk-stellar mass scaling relation in nearby regions}\label{sect:comparison}
The four nearby regions of Taurus ($d=140\,pc$, age$\sim$1-2\,Myr, \citealt{luhman04}), Lupus ($d=140\,pc$, age$\sim$1-3\,Myr, \citealt{comeron08}), \ChaI{} ($d=160\,pc$, age$\sim$2-3\,Myr, \citealt{luhman08}), and Upper Sco ($d=145$, age$\sim$5-10\,Myr, \citealt{slesnick08}) have ages spanning the range over which significant disk evolution is expected to occur, hence they have been the focus of many studies to understand when and how protoplanetary material is dispersed. Infrared surveys with the Spitzer Space Telescope have established that the fraction of optically thick dust disks, those displaying excess emission at IRAC wavelengths (3.6-4.5\,\micron), decreases from $\sim$65\% in Taurus to $\sim 50$\% in Lupus and \ChaI{} and drops to only $\sim$15\% in Upper Sco  \citep{ribas14}. Over the same age range there is tentative evidence for  an increase in the frequency of Class II/T SEDs relative to the total disk population, just a few \% at ages $\le$2\,Myr and $\sim$10\% at older ages \citep{espaillat14}. These observations trace the depletion/dispersal of small micron-sized grains within a few AU from the star and  support a scenario in which protoplanetary material is cleared from inside out (see \citealt{alexander14} for a recent review on disk dispersal timescales and mechanisms). Millimeter observations probe the population of larger mm/cm sized-grains at radial distances $\ga 10$\,AU. Thanks to the exquisite sensitivity of ALMA there are now millimeter surveys that parallel those at infrared wavelengths in sample size, thus enabling testing if significant evolution occurs in the outer disk over the $\sim$1 to 10\,Myr age range.

The disk populations of the \ChaI{} (this paper), Lupus \citep{ansdell16}, and Upper~Sco \citep{bare16} regions have been probed with ALMA in Band~7 at similar sensitivity. The Taurus star-forming region has been covered with the SMA at a lower sensitivity \citep{andrews13}, about 3 and 15 times lower than that used here for the {\it Hot} and {\it Cool} samples, respectively. To compare their $M_{\rm dust} - M_*$ relations we re-analyze all the datasets in a self-consistent manner: we re-compute all the stellar masses as discussed in Section~\ref{sect:stellar_masses} using the same evolutionary tracks and then apply the approach described in Section~\ref{sect:DustDisk} to account for mm detections and upper limits. The first step is important because, as pointed out in \citet{andrews13}, different evolutionary tracks can result in slightly different $M_{\rm dust} - M_*$ relations.  We note that the adopted spectral type-effective temperature scale is essentially the same in all 4 regions with a small difference of only $\sim$10\,K in the M7-M8 range where there are only a few, if any, sources in each region.
For Upper~Sco we only consider disks classified as 'Full' and 'Transitional' in Table~1 of \citet{bare16}, equivalent to the Class~II and II/T SEDs in \ChaI. More evolved/debris disks, Class~III-type, are not included in the Taurus, Lupus, and \ChaI{} millimeter surveys. These disks most likely represent a different evolutionary stage when most of the gas disk has been dispersed (e.g. \citealt{pascucci06}) and the millimeter emission arises from second generation dust produced in the collision of larger asteroid-size bodies. 
The resulting $M_{\rm dust} - M_*$ relations for these four regions are summarized in Table~\ref{tab:relations} and plotted in Figure~\ref{fig:comp_SF} for the case of constant dust temperature. This case  is essentially equivalent to comparing sub-millimeter luminosities as a function of $M_*$ in different star-forming regions (see also Section~\ref{sect:DustDisk}).
 
\begin{figure*}
\centering
\includegraphics[width=1\textwidth]{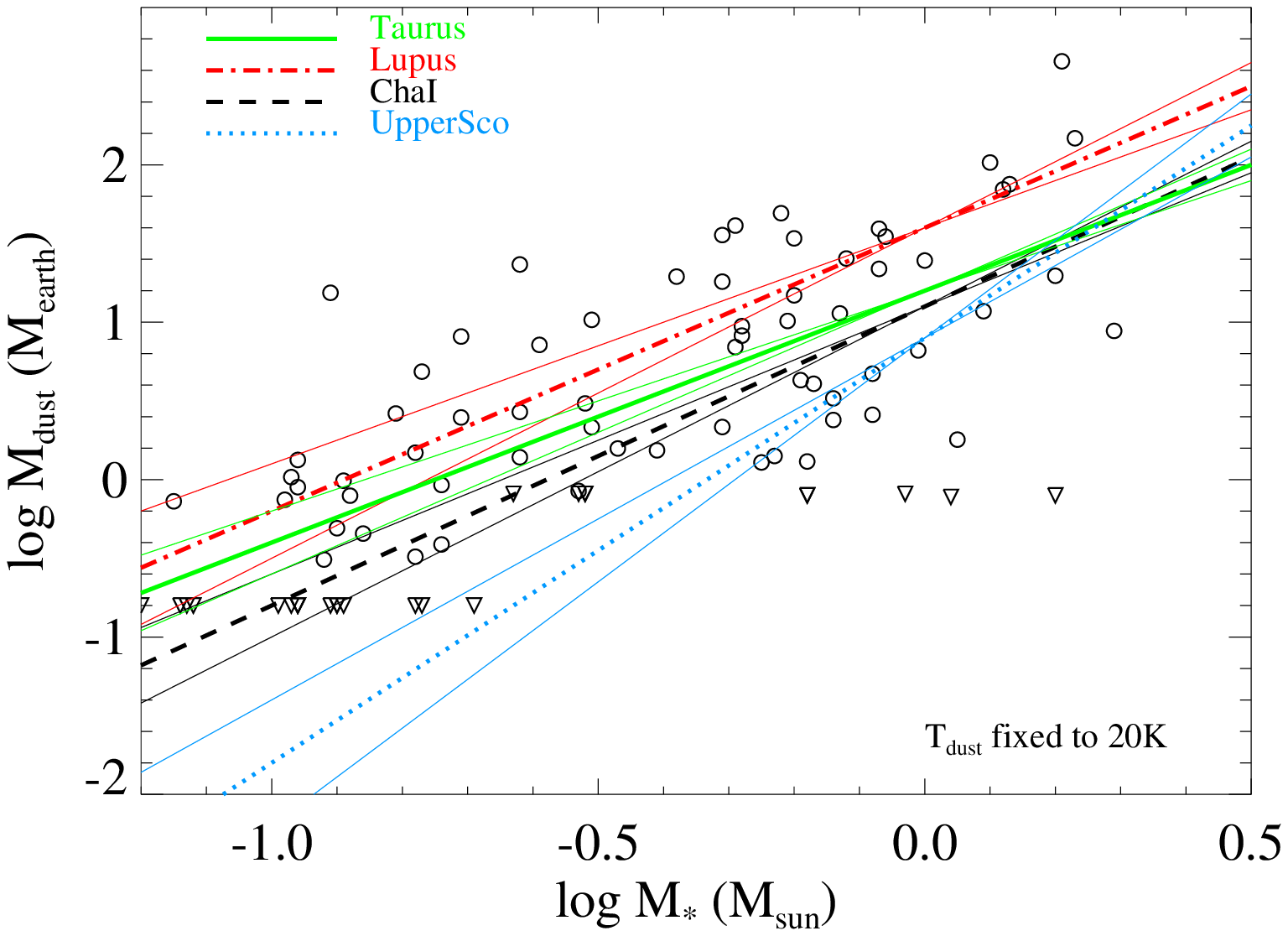}
\caption{The $M_{\rm dust}-M_*$ relation in four different regions:
Taurus (green solid line), Lupus (red dot-dashed line), \ChaI{} (black dashed line), and Upper~Sco (light blue dotted line). 
These relations are obtained assuming a fixed dust temperature of 20\,K (see also Table~\ref{tab:relations}).
For \ChaI{} we also plot the individual dust disk masses. Note that the $\sim$10\,Myr-old Upper~Sco has a steeper  $M_{\rm dust}-M_*$ relation than the other star-forming regions.
} 
\label{fig:comp_SF}
\end{figure*}
 
Taurus has the shallowest  $M_{\rm dust} - M_*$ relation among these regions. However, as discussed in Section~\ref{sect:results}, the lower sensitivity of the survey can account for the apparent difference with \ChaI{}. Lupus has the same slope but appears to have slightly more massive disks than Taurus and \ChaI{}. However, given the few $\sim 1$\,$M_\sun$ stars in Lupus the intercept is less well determined than in Taurus and \ChaI . Indeed, adding the 20 obscured Lupus sources by randomly assigning a stellar mass reduces the intercept by 0.3 making the $M_{\rm dust} - M_*$ relation of Lupus the same as the one of Taurus (Ansdell private communication) and \ChaI. Hence, we conclude that the same $M_{\rm dust} - M_*$ relation is shared by star-forming regions that are 1-3\,Myr old. We also note that the relation is steeper than linear. As already pointed out in \citet{bare16} and \citet{ansdell16} the disk mass distribution in the $\sim$5-10\,Myr-old Upper~Sco association is significantly different from that in Taurus and Lupus, with the mean dust disk mass in the latter two regions being about 3 times higher than in Upper~Sco. By performing a generalized Wilcoxon test\footnote{The null hypothesis is that two groups have the same distribution, p denotes the probability to reject the null hypothesis. Censored data are included in cendiff.} with the cendiff command in the NADA R package, we find that the disk mass distribution in \ChaI{} is indistinguishable from that of Taurus (p=52\%) and Lupus (p=8\%) but different from that of Upper~Sco (p=0.0001\%) within the same $\sim 0.1-1.6\,M_\sun$ stellar-mass range. The mean dust disk mass is $\sim 10\,M_\earth$ for \ChaI{} but only $\sim 4\,M_\earth$ for Upper~Sco in the assumption of constant dust temperature and with our value for the dust opacity.} Table~\ref{tab:relations} shows that the 
$M_{\rm dust} - M_*$ relation is also steeper in Upper~Sco than in the other three younger regions (see also Fig.~6 in \citealt{ansdell16}). Based on the inferred relations, it appears that disks around 0.5\,$M_\sun$ have depleted their dust disk mass in mm grains by a factor of 2.5 by $\sim$10\,Myr, while disks around 0.1\,$M_\sun$ by an even larger factor of 5. To further corroborate our finding, we perform the same Wilcoxon test on the disk mass distribution for stars more and less massive than $\sim 0.5\,M_\sun$. The probability that \ChaI{} and Upper~Sco have the same disk mass distribution is as high as 52\% for $>0.5\,M_\sun$ stars while it is only 0.02\% for the lower stellar mass bin with average masses that are a factor of 2 lower in Upper~Sco than in \ChaI . With lowering the stellar mass value to create the two disk mass samples, the probability that the high-stellar mass bin in \ChaI{} and Upper~Sco have the same disk mass distribution also decreases reaching 1\% at $0.28\,M_\sun$.
This demonstrates that differences in the two distributions are more pronounced toward the lower stellar mass end, well in line with a steeper $M_{\rm dust} - M_*$ relation in Upper~Sco than in \ChaI.

Finally, it is interesting to note that the dispersion around the  $M_{\rm dust} - M_*$ relations is very similar in the four regions and amounts to $\sim$0.8\,dex. Different disk masses, dust temperatures, and grain sizes can contribute to the dispersion. Whatever the cause, the dispersion does not depend on the environment or age of the region, but seems to be an intrinsic property of the disk population reflecting a range of initial conditions which might, at least in part, account for the diversity of planetary systems.

\subsection{On the evolving disk-stellar mass scaling relation}
In the previous Section we showed that the 1-3\,Myr-old star-forming regions of Taurus, Lupus, and \ChaI{} share the same $M_{\rm dust} - M_*$ relation while the older Upper~Sco association has a steeper relation. What is the physical process leading to a steepening of the $M_{\rm dust} - M_*$ relation with time?
One possibility would be to invoke a stellar-mass dependent conversion to larger grains, in that disks around lower mass stars  would convert more mm grains into larger cm grains that go undetected. Alternatively, the higher depletion of mm-sized grains toward lower-mass stars could result from more  efficient inward drift, i.e. mm-sized grains 
would be still orbiting the star but in the inner and not in the outer disk where optical depth effects might hide them.

To test these scenarios we use the Lagrangian code developed by \citet{krijt16} and simulate the evolution of dust disk grains subject to: a) growth and fragmentation; b) growth and radial drift; and c)  growth, radial drift, and fragmentation. In all models, the dust disk initially extends from 2 to 200\,AU with a power-law surface density with index -1.5, the dust-to-gas mass ratio is 0.01, the total mass equals 1\% of the central star mass, the turbulence is characterized by $\alpha=0.01$ \citep{ss73}, the fragmentation velocity is 3\,m/s, and the grain porosity is constant at 30\%. The code calculates the radial profile of the mass-dominating grain size and the dust surface density, which is then integrated to obtain the total mm flux as a function of time. 

\begin{figure}[h]
\centering
\includegraphics[width=0.5\textwidth]{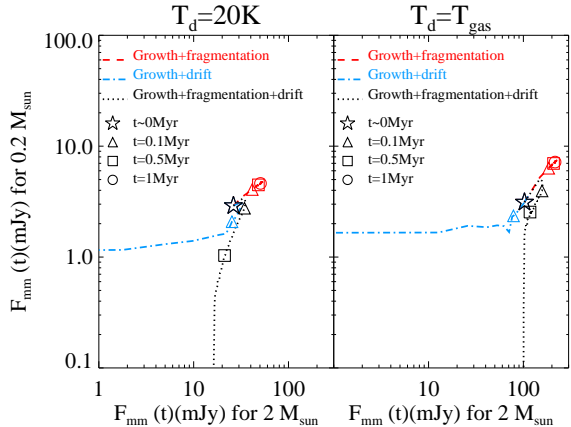}
\caption{Evolution of the 850\,\micron{} flux density for a disk around a 2\,$M_\sun$ star (x-axis) and a disk around a 0.2\,$M_\sun$ star (y-axis).
For both stars, the initial disk mass is set equal to 1\%{} of the stellar mass. The left panel assumes a constant dust temperature through the disk while the right panel a radially decreasing dust temperature set equal to the gas temperature, see text for details. The growth+fragmentation+drift case can qualitatively explain the observed steepening in the $M_{\rm dust}-M_*$ relation with time.
} 
\label{fig:models}
\end{figure}

Figure~\ref{fig:models}  shows the evolution of the mm flux around two stars, one having a mass equal to 0.2\,$M_\sun$ (y-axis) and the other  2\,$M_\sun$ (x-axis). In the left panel the dust disk temperature is assumed to be fixed to 20\,K while in the right panel it varies radially and equals the gas disk temperature which is prescribed to decrease with radius and be higher around high-mass stars:  $T_{\rm gas}=280K \times (r/AU)^{-0.5} (M_*/M_\sun)^{0.5}$. While the resulting mm fluxes depend on the assumed dust disk temperature, as highlighted in Section~\ref{sect:DustDisk}, the evolutionary behavior is the same. More specifically, growth and fragmentation (red dashed line and symbols) do not change the initial flux ratio of the two disks, hence cannot explain the steepening of the $M_{\rm dust} - M_*$ relation with time. Growth and drift (light blue dot-dashed line and symbols) are faster in denser disks around higher-mass stars, hence these disks are depleted faster of mm grains and become mm faint sooner than disks around lower mass stars. This is opposite to what is observed. Finally, the more realistic case of growth, radial drift, and fragmentation (black dotted line and symbols) shows a behavior consistent with the observations, in that the disk around the 0.2\,$M_\sun$ reduces its mm flux faster than the disk around the 2\,$M_\sun$ star. This is because the timescale on which radial drift removes the largest grains is shorter around low-mass stars. As a result, the disk around the 2\,$M_\sun$ star can remain mm bright longer.
To first order the timescale over which dust is removed is the inverse of the Stokes number ($St$) of the largest grains which, in the Epstein regime, scales as $St^{-1} \propto \alpha (c_s / v_{frag})^2 \propto (M_*)^{0.5}$ in the fragmentation-limited case and as $St^{-1} \propto (c_s / v_K)^2 \propto (M_*)^{-0.5}$ in the drift-limited case \citep{birnstiel12}. Thus, dust removal is faster around lower-mass stars only in the fragmentation-limited case.
For the specific models shown in Figure~\ref{fig:models}, the maximum grain size is $<$0.1\,mm outside of $\sim$50\,AU around the 2\,$M_\sun$ star and outside of $\sim$15\,AU around the 0.2\,$M_\sun$ star.

In summary, the comparison between models and observations suggests that the maximum grain size in the outer disk is fragmentation-limited, rather than drift-limited. 
 As already pointed out in the literature (e.g. \citealt{pinilla13}), a reduced drift efficiency, perhaps caused by radial pressure bumps,  is necessary in all models to match the observed lifetime of disks at mm wavelengths.

%

The scenarios discussed above can be tested with future millimeter observations. 
If grain growth from mm to cm in size is responsible for the steepening of the $M_{\rm dust} - M_*$ relation with time \citep{bare16}, we should expect a stellar-mass- and time-dependent power law index $\beta$ of the dust opacity. More specifically, older disks around lower-mass stars should have a lower $\beta$ than disks around younger higher-mass stars.
A dependence of $\beta$ with stellar mass is not seen for Taurus disks around $\sim 0.4-2.2\,M_\sun$ stars and for a few disks around sub-stellar objects \citep{ricci10,ricci14}, but it should be tested if it arises over a statistically significant sample of disks spanning a broad range in stellar masses and at later evolutionary times. If instead the maximum grain size is fragmentation limited as we suggest, the $\beta$ dependence with stellar mass would be opposite because higher-mass stars would have, on average, larger grains in their disks than lower-mass stars. In addition, there would not be a time-dependence because the fragmentation-limited regime is insensitive to the surface density evolution \citep{birnstiel12}.  

Another prediction of this scenario is that disks around lower-mass stars would be smaller in size than disks around higher-mass stars. Previous work has pointed out that dust disk radii correlate positively with mm fluxes for T~Tauri stars \citep{isella09,andrews10,guilloteau11} but the scatter is large and what is really needed is to demonstrate a correlation with stellar mass. In the sub-stellar regime, there are only five disks whose dust disk radii at mm wavelengths can be reliably inferred. The three in Taurus are rather large ($50-100$\,AU, \citealt{ricci14}) while the two in $\rho~Oph$ are much smaller ($<25$\,AU, \citealt{testi16}).  A systematic ALMA survey with high spatial resolution and sufficient S/N is missing.

Finally, we would like to comment on the finding of a longer disk lifetime around low-mass stars based on infrared observations. 
\citet{carpenter06} found that the fraction of optically thick disks in Upper~Sco is higher for K+M dwarfs ($\sim 0.1-1\,M_\sun$) than for  earlier spectral type stars. Expanding upon this, \citet{bayo12}  reported a higher fraction of optical thick disks around stars less massive than $\sim 0.6\,M_\sun$ in the 5-12\,Myr-old Collinder~69 cluster. These results demonstrate that the inner disk of low-mass stars is not depleted of micron-sized grains but do not place any constraint on the outer disk. On the contrary, the ALMA observations presented in this paper trace the population of mm  grains in the outer disk. Inward drifting mm grains that collide and replenish the inner disk of smaller sub-micron grains might explain both the apparent lack of mm grains in the outer disk and the longer lived
optically thick disks around low-mass stars.

\subsection{The mass accretion rate-disk mass relation}
In the classical paradigm of disk evolution, the accretion of disk gas onto the star is thought to result from the
coupling of the stellar magnetic field with ions in so-called active layers of the
disk (magneto-rotational instability model, e.g., \citealt{gammie1996}). However, in this
standard picture the accretion rate is independent from the mass of the central
star. \citet{hartmann06} showed that a weak linear dependence can be
recovered when including stellar irradiation as a disk heating mechanism in
addition to viscous accretion. Further steepening the relation would be possible if
disks around very low-mass stars are less massive, fully magnetically active, and as such having
viscously evolved substantially \citep{hartmann06}.
Alternatively, \citet{ercolano14} 
have proposed that the $\dot{M}-M_*$ relation is flatter for spectral
types earlier than M due to a specific disk dispersal mechanism, star-driven X-ray
photoevaporation. 
Looking at the complete stellar mass range,  \citet{dullemond06} have shown that a steep $\dot{M} \sim (M_*)^{1.8}$ relation 
arises naturally if the centrifugal radius of the parent core is independent of the mass of the core
and the spread in $\dot{M}$ at any stellar mass would reflect an initial distribution of core rotation rates.
In all cases, $\dot{M}$ should scale linearly with the disk mass, implying that
the $\dot{M}-M_*$ relation should be the same as the disk mass-stellar mass relation.

The $\dot{M}-M_*$ relation has been determined for Taurus, Lupus, and \ChaI{} while it is not available for Upper~Sco. For these three young regions the relation is close to a power law of two: $\dot{M} \propto (M_*)^{1.9\pm0.3}$ for Taurus \citep{hh08}; $\dot{M} \propto (M_*)^{1.8\pm0.2}$ for Lupus \citep{alcala14}; and  $\dot{M} \propto (M_*)^{1.7\pm0.4}$ for \ChaI{} \citep{manara16a}. 
While we do not have total (gas+dust) disk masses, it is interesting to note that $M_{\rm dust}$ displays the same steep relation with $M_*$ in these three regions if the average dust temperature is constant, while the relation is slightly shallower for a dust temperature scaling with stellar luminosity (see Table~\ref{tab:relations}). A more robust way to test the basic prediction of a linear relation between $\dot{M}$ and disk mass is to directly relate these quantities for the same large sample of objects belonging to the same star-forming region. This could be recently achieved for the Lupus clouds.
Assuming a constant dust temperature  to convert millimeter fluxes into dust disk masses, \citet{manara16b} showed that $\dot{M}$ and $M_{\rm dust}$ are correlated in Lupus in a way that is compatible with viscous evolution models. Interestingly, the gas disk mass inferred from CO isotopologues does not show a similar correlation with $\dot{M}$. This may be the result of CO not being a good tracer of the total gas disk mass because carbon can be sequestered in more complex molecules on icy grains (e.g. \citealt{bergin14}) and/or because of complex isotope-selective processes \citep{miotello14,miotello16}. It would be interesting to extend such studies to other regions, especially Upper~Sco, where the $M_{\rm dust} - M_*$ relation is even steeper than in younger star-forming regions.

\subsection{Total disk masses and planetary systems}\label{sect:Kepler}
Given the relevance of disk masses to planet formation models, we discuss here the uncertainties in estimating total disk masses, whether disks appear to be close to being gravitationally unstable, and how dust disk masses compare to the amount of solids  locked into exoplanets.

As discussed in Section~\ref{sect:DustDisk} the average disk temperature tracing mm emission affects the absolute value of the dust disk mass, as well as the disk-stellar mass scaling relation, with cooler temperatures leading to higher disk mass estimates. For the two temperature relations adopted here the average difference in dust disk masses  amounts to a factor of $\sim$3.  An even larger uncertainty is introduced by the dust opacity which depends on grain composition as well as size distribution (see e.g. \citealt{testi14}), which are both still poorly constrained.  Silicates constitute the main source of opacity at $\sim$1\,mm. While plausible uncertainties in their optical constants affect the dust opacity by no more than a factor of two, porosity adds an uncertainty of a factor of several for grains larger than 100\,\micron{} \citep{pollack94,hs96,semenov03}. Even assuming a fixed dust composition, the 1.3\,mm opacity can vary by a factor of $\sim$4 depending on whether the grain size distribution extends to 1\,cm (low opacity=higher mass) or to 0.8\,mm (high opacity=lower mass), see e.g. Figure~1 in \citet{tazzari16}.
The 2.3\,cm$^2/$g dust opacity we have adopted is close to the one for a grain size distribution extending to 1\,cm. This means that if the true grain size distribution were truncated at 1\,mm the dust disk masses would be a factor of 4 lower than those we report. Given that our choice of dust opacity maximizes dust disk masses over the range of grain sizes expected/detectable in the outer disk, we will continue our discussion adopting the dust disk masses obtained with a constant dust temperature which, instead, minimizes the disk masses toward lower-mass stars.
  
\begin{figure}[h]
\centering
\includegraphics[width=0.5\textwidth]{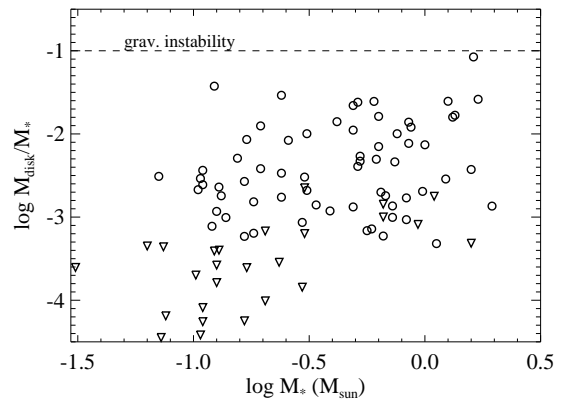}
\caption{Disk to stellar mass ratios as a function of stellar masses ($M_*$) for our \ChaI{} region. Disk masses are dust masses (using a constant dust disk temperature of 20\,K) multiplied by 100.
} 
\label{fig:MdiskoMstar_vs_Mstar}
\end{figure}

Figure~\ref{fig:MdiskoMstar_vs_Mstar} shows the distribution of $M_{\rm disk}/M_{*}$ where $M_{\rm disk}$ is simply the dust disk mass multiplied by the ISM gas-to-dust ratio of 100. Although recent gas mass estimates using rotational lines from CO isotopologues have claimed gas-to-dust ratios well below the ISM value in young disks \citep{wb14,ansdell16}, detailed physico-chemical disk models need to be carried out to properly account for isotope-selective processes \citep{miotello14,miotello16}. In addition, carbon can be extracted from CO via reactions with He$^+$ and form hydrocarbons that freeze-out, thus reducing the CO abundance in the disk atmosphere \citep{favre13}.  Indeed, the only disk with an independent mass estimate, using the HD (J=1-0) transition at far-infrared wavelengths, has a gas-to-dust mass ratio consistent with the ISM value and confirms that masses using CO isotopologues can be off by up to a factor of 100 \citep{bergin13}. With these caveats it is interesting to compare the inferred distribution of $M_{\rm disk}/M_{*}$ ratios to the limiting mass ratio above which gravitational instabilities set in ($M_{\rm disk}/M_{*} \sim 0.1$, e.g., \citealt{lodato05}, dashed line in Figure~\ref{fig:MdiskoMstar_vs_Mstar}). While the median $M_{\rm disk}/M_{*}$ value of $\sim 0.04$ is well below 0.1, the brightest source in our sample is close to the gravitational instability boundary. In addition, six other sources, ranging in stellar mass from $\sim 0.15$ to 1.7\,$M_{\sun}$,  have ratios only  a factor of 4 lower than the  gravitational instability limit and appear to delineate an upper horizontal boundary. 
It is interesting to speculate that this upper boundary is the one set by gravitational instability, but independent observations of the gas content are necessary to make any firm conclusion.

\begin{figure}[h]
\centering
\includegraphics[width=0.5\textwidth]{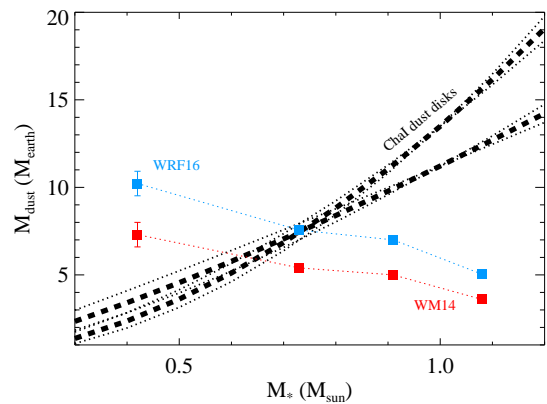}
\caption{Dust disk masses in \ChaI{} (black dashed and dotted lines) compared to the average mass in solids (red and blue squares) from the {\it Kepler} exoplanets as computed in \citet{mulders15b}. The planet mass-radius relation by WRF16 \citep{wrf15} gives a higher average mass than that by WM14 \citep{wm14}. Regardless of the assumed relation, low-mass stars ($\sim 0.4 M_\odot$) have dust disk masses  lower than the average mass locked up in close-in exoplanets.  
} 
\label{fig:Kepler}
\end{figure}

How do disk masses compare with the mass locked up in exoplanets around other stars? \citet{nk14} used a Monte Carlo approach to create ensembles of systems with planets and debris disks at their known incidence rates and compared them to the Taurus protoplanetary disk masses from \citet{andrews13}. They found that the mass in solids in Class~II sources are barely enough to account for the known population of {\it Kepler} and RV planets plus debris disks and seems to fall short for the 5-30\,$M_\earth$ planets at 0.5-10\,AU discovered by microlensing. \citet{mulders15b} focused on stellar mass dependencies in the amount of solids from the well-characterized {\it Kepler} survey, probing planets with periods within 50\,days ($\sim$0.3\,AU around a solar mass star). They pointed out that the average  mass in solids locked up in exoplanets increases roughly inversely with stellar mass instead of decreasing as the dust disk mass estimated from millimeter observations. Figure~\ref{fig:Kepler} compares dust disk masses in \ChaI{} with the solid mass in exoplanets. For solar or higher-mass stars dust disk masses are larger than the mass of solids locked up in close-in exoplanets. However, $\sim$2\,Myr-old disks around low-mass stars ($\sim 0.4 M_\odot$) appear to be already short in solids by a factor of at least 2 to reproduce the average mass in exoplanets. At $\sim10$\,Myr the deficit amounts to more than a factor of 5 as shown by the Upper~Sco region. Recently, \citet{gillon16} reported the discovery of 3 close-in ($<0.1$\,AU) Earth-size planets around the 0.08\,$M_\sun$ star TRAPPIST-1. Interestingly, the largest dust disk mass that we can obtain from the relations in Table~4 for such a star is only 1.6$M_\earth$, not enough to reproduce the total mass in the TRAPPIST-1 planetary system. Even if half of the disk mass is already converted into planetesimals in $\sim 1$\,Myr-old disks as proposed by \citet{nk14}, dust disk masses around low-mass stars are still on the low side to account for the solid mass in close-in exoplanets. As discussed in Section~\ref{sect:comparison}, inward drift most likely 
contributes to redistribute the mass of millimeter grains early on. If so,  there should be a large population of millimeter grains closer in to the star at radii where our observations are not sensitive to. It is unclear if such grains will retain their size for long or quickly grow to form the close-in planets we see around mature stars.

\section{Conclusions}
We presented an ALMA 887\,\micron{} survey of the disk population around objects from $\sim 2$ to 0.03\,$M_\sun$ in 
the nearby $\sim$2\,Myr-old \ChaI{} star-forming region. 
One of our main goals was to use the continuum emission to estimate dust disk masses and establish how they scale with stellar mass. Our main findings can be summarized as follows:
\begin{itemize}
\item We detect thermal dust emission from 66 out of 93 disks, spatially resolve 34 of them, and identify two disks with large dust cavities ($\sim$45\,AU in radius). 
\item We find that the disk-stellar mass scaling relation in \ChaI{} is steeper than linear: $M_{\rm dust} \propto (M_*)^{1.3-1.9}$, where the range in the power law index reflects two extreme relations between the average dust temperature and stellar luminosity. 
\item By re-analyzing in a self-consistent way all millimeter data available for nearby regions, we show that the 1-3\,Myr-old regions of Taurus, Lupus, and \ChaI{} have the same $M_{\rm dust}-M_*$ relation  while the 10\,Myr-old Upper~Sco association has an even steeper relation.
\item The dispersion around the $M_{\rm dust}-M_*$ relation is very similar among regions with ages $\sim1-10$\,Myr hinting at a range of initial conditions which might partly account for the diversity of planetary systems.
\item The slopes of the $M_{\rm dust}-M_*$ and of the  $\dot{M}-M_*$ relations are the same for Taurus, \ChaI, and Lupus when assuming a constant dust temperature, in agreement with the basic expectation from viscous disk models. 
\end{itemize}
By comparing our results with theoretical models of grain growth, drift, and fragmentation we show that a steeping of the $M_{\rm dust}-M_*$ relation with time occurs if outer disks are in the fragmentation-limited regime. 
This is because when fragmentation sets the largest grain size, radial drift will occur at shorter timescales around lower-mass stars.
This scenario of redistributing mass in the disk can also account for the apparent lack of solids in Myr-old disks around low-mass stars ($\le 0.4\,M_\sun$) when compared to the average mass of solids locked into close-in exoplanets. Such a scenario results in a stellar-mass-dependent but not a time-dependent power-law index of the dust opacity. It also implies a stellar-mass-dependent disk size for mm grains.  
Deeper and higher resolution millimeter observations are needed to test the predicted trends. Establishing if and how the size of dust disks scales with stellar mass will also enable to measure the dependence between the  average dust temperature and stellar luminosity which is crucial to pin down the exact $M_{\rm dust}-M_*$ relation. 


\acknowledgments
The authors thank the anonymous referee and the statistic editor for insightful comments that helped improving the manuscript.
IP thanks Megan Ansdell and John Carpenter for sharing some of their results in advance of publication and for clarifying their procedures to analyze the ALMA data.  IP also acknowledges support from a NSF Astronomy \& Astrophysics Research Grant (ID: 1515392). GH and LF are supported by general grant 11473005 awarded by the National Science Foundation of China.
CFM gratefully acknowledges an ESA Research Fellowship. LT acknowledges partial support from Italian Ministero dell\textsc{\char13}Istruzione, Universit\`a e Ricerca through the grant Progetti Premiali 2012 -- iALMA (CUP C52I13000140001) and  from Gothenburg Centre of Advanced Studies in Science and Technology through the program {\it Origins of habitable planets}. 
This material is based upon work supported by the National Aeronautics and Space Administration under Agreement No. NNX15AD94G for the program {\it Earths in Other Solar Systems}. The results reported herein benefitted from collaborations and/or information exchange within NASA\textsc{\char13}s Nexus for Exoplanet System Science (NExSS) research coordination network sponsored by NASA\textsc{\char13}s Science Mission Directorate.

{\it Facilities:} \facility{ALMA}.



\appendix
\section{Comparison of linear regression methods}\label{sect:regression}

Here, we compare different linear regression methods to fit the  $F_{\rm mm}-M_*$ relation in the log-log plane.
We will show how the intrinsic scatter in the relation and censored values 
(upper limits to the millimeter flux density) contribute to the best fit slope and intercept.

We start by comparing the results from two IDL routines (fitexy and mpfitexy) that do not account for upper limits,
i.e. we only fit the 66 sources with measured flux densities. Both routines assume symmetric measurement errors 
in x [log($M_*$)] and y [log($F_{\rm mm}$)] and use the Nukers' estimator to find the best fit (see e.g. \citealt{tremaine02}).
The main difference is that mpfitexy accounts for the intrinsic scatter and can automatically adjust it to ensure a reduced $\chi^2$ of unity.
Indeed, this is necessary for our dataset where $F_{\rm mm}$ has a large spread at each stellar mass and confirmed
by the fact that fitexy cannot find a good fit, the $\chi^2$ is greater than 550 and the probability that the model is correct is zero. 
The mpfitexy requires a scatter about the relation of 0.5\,dex to obtain $\chi^2 \sim 1$. 
In addition, the uncertainties in the slope and intercept from fitexy are unrealistically low and the best fit is dominated by a few precise measurements 
when not accounting for the scatter thus biasing the derived slope and intercept (see Table~\ref{tab:tests}).
These issues are well documented in \citet{tremaine02}.

Next, we compare three different routines that account for censored data using different methods.
The one utilized throughout the paper is the linmix\_err routine (IDL version) written by \citet{kelly07}
and already used in several other astronomical applications. 
This routine accounts for both measurement errors and intrinsic scatter while the other two routines 
from the R statistical package (censReg and cenken) do not include individual measurement errors. The fact that they all provide
the same slope and intercept within the quoted uncertainties (Table~\ref{tab:tests}) again confirms that the intrinsic scatter about the relation drives the best fit.
In what follows we briefly summarize the methods used in these routines and additional lessons learned from the comparison.

The linmix\_err routine uses a Bayesian approach assuming a normal linear regression model, i.e.
the conditional distribution is a normal density, and computes the likelihood function of the data by integrating the
conditional distribution. The measurement errors and the intrinsic scatter about the line are all assumed to be normally distributed. 
A Markov chain Monte Carlo method is used to compute the uncertainties on the slope and intercept.
Further details about the approach are summarized in \citet{kelly07}.

The censReg R routine is based on the parametric Maximum likelihood estimation and assumes a normal distribution
of the error term (see e.g. \citealt{greene08}). As mentioned above individual measurement errors on x and y are not taken into account and
one single left censoring (upper limit) is considered.
Because our survey has different upper limits for the {\it Hot}  and {\it Cool} samples, we had to use the less stringent one, the one from the {\it Hot} sample.
In other words, the results reported in Table~\ref{tab:tests} are from treating 58 datapoints as uncensored (detections) and 35 as upper limits,
27 of them are true upper limits while 8 are additional detections below the upper limit set by the {\it Hot} sample.

Finally, the cenken R routine uses the non-parametric Akritas-Thiel-Sen line with the Turnbull estimate of intercept \citep{akritas95}.
The advantage of this method is that it does not make any assumption about the distribution of the data.
While measurement errors on x and y are not included, upper limits can be specified individually meaning that both the {\it Hot}
 and {\it Cool} sample upper limits can be properly taken into account.
 
As summarized in Table~\ref{tab:tests} the three routines treating censored data find the same slope and intercept for the 
$F_{\rm mm}-M_*$ relation. As expected, the slope is steeper and the intercept is lower than that obtained considering only uncensored data 
but properly accounting for the scatter (mpfitexy).
The slightly lower slope from censReg probably reflects that the {\it Cool} sample upper limits are not treated  (see also Section~\ref{sect:results}
for a similar effect when applying an even shallower cutoff as in the Taurus survey). 
Finally, the fact that parametric and non-parametric approaches reach the same results suggest that
the  slope and intercept of the  $F_{\rm mm}-M_*$ relation are not affected by the underlining assumptions on the distribution of the data.

\clearpage

\clearpage

\begin{deluxetable}{lccccccccc}
\tabletypesize{\tiny}
\tablecaption{Source Properties\label{tab:SourceProperties}}
\tablewidth{0pt}
\tablehead{
\colhead{2MASS} & \colhead{Other} & \colhead{Multiplicity} &  \colhead{Ref.} & \colhead{SpTy} & \colhead{SED} & \colhead{ALMA} &  \colhead{SpTy} &  \colhead{Ref.} &  \colhead{log($M_*$)}\\ 
\colhead{ } & \colhead{Name} & \colhead{(\arcsec)} & \colhead{} & \colhead{Luhman} &  \colhead{}  &  \colhead{Sample}  &  \colhead{adopted} &  \colhead{} & \colhead{($M_\sun$)}
}
\startdata
J10533978-7712338 &              &       &      & M2.75 &    II &   Hot &    M2 & M16b  & -0.41\tablenotemark{\dagger}\\ 
J10555973-7724399 &           T3 & 2.210 &  D13 &    M0 &    II &   Hot &    K7 & M16a  & -0.13 (-0.17,-0.07) \\ 
J10561638-7630530 &     ESOH$\alpha$553 &       &      &  M5.6 &    II &  Cool &  M6.5 & M16b  & -0.96 (-1.03,-0.89) \\ 
J10563044-7711393 &           T4 &       &      &  M0.5 &    II &   Hot &    K7 & M16a  & -0.07 (-0.15, 0.19) \\ 
J10574219-7659356 &           T5 & 0.160 &  N12 & M3.25 &    II &   Hot &    M3 & M16b  & -0.52 (-0.57,-0.47) \\ 
J10580597-7711501 &              &       &      & M5.25 &    II &  Cool &  M5.5 & M16b  & -0.96 (-1.06,-0.86) \\ 
J10581677-7717170 &        SzCha & 5.120 &  D13 &    K0 &  II/T &   Hot &    K2 &  M14  &  0.10 ( 0.06, 0.14) \\ 
J10590108-7722407 &        TWCha &       &      &    K2 &    II &   Hot &    K7 & M16a  & -0.07 (-0.14, 0.17) \\ 
J10590699-7701404 &        CRCha &       &      &    K2 &    II &   Hot &    K0 & M16a  &  0.23 ( 0.18, 0.28) \\ 
J11004022-7619280 &          T10 &       &      & M3.75 &    II &  Cool &    M4 & M16b  & -0.62 (-0.69,-0.54) \\ 
J11022491-7733357 &        CSCha &       &      &    K6 &  II/T &   Hot &    K2 &  M14  &  0.13 ( 0.09, 0.20) \\ 
J11023265-7729129 &       CHXR71 & 0.560 &  D13 &    M3 &    II &   Hot &    M3 & M16b  & -0.52 (-0.58,-0.45) \\ 
J11025504-7721508 &          T12 &       &      &  M4.5 &    II &  Cool &  M4.5 & M16a  & -0.74 (-1.23,-0.68) \\ 
J11040425-7639328 &     CHSM1715 &       &      & M4.25 &    II &  Cool &  M4.5 & M16b  & -0.74 (-0.83,-0.65) \\ 
J11040909-7627193 &       CTChaA & 2.670 &  D13 &    K5 &    II &   Hot &    K5 & M16a  & -0.06 (-0.16, 0.04) \\ 
J11044258-7741571 &        ISO52 &       &      &    M4 &    II &  Cool &    M4 & M16a  & -0.62 (-0.69,-0.54) \\ 
J11045701-7715569 &          T16 &       &      &    M3 &    II &   Hot &    M3 & M16b  & -0.53 (-0.59,-0.47) \\ 
J11062554-7633418 &     ESOH$\alpha$559 &       &      & M5.25 &    II &  Cool &  M5.5 & M16b  & -0.91 (-1.01,-0.81) \\ 
J11062942-7724586 &              &       &      &    M6 &    II &  Cool &    M6 &  L07  & -1.12 (-1.66,-1.00) \\ 
J11063276-7625210 &     CHSM7869 &       &      &    M6 &    II &  Cool &  M6.5 & M16b  & -1.13 (-1.25,-0.97) \\ 
J11063945-7736052 &        ISO79 &       &      & M5.25 &    II &  Cool &    M5 & M16b  & -0.78\tablenotemark{\dagger} \\ 
J11064180-7635489 &          Hn5 &       &      &  M4.5 &    II &  Cool &    M5 & M16a  & -0.78 (-0.86,-0.68) \\ 
J11064510-7727023 &       CHXR20 & 28.46 & KH07 &    K6 &    II &   Hot &    K6 & M16b  & -0.03 (-0.10, 0.22) \\ 
J11065906-7718535 &          T23 &       &      & M4.25 &    II &  Cool &  M4.5 & M16a  & -0.71\tablenotemark{\dagger}  \\ 
J11065939-7530559 &              &       &      & M5.25 &    II &  Cool &  M5.5 & M16b  & -0.97 (-1.07,-0.87) \\ 
J11070925-7718471 &              &       &      &    M3 &    II &   Hot &    M3 &  L07  & -0.52 (-0.58,-0.45) \\ 
J11071181-7625501 &     CHSM9484 &       &      & M5.25 &    II &  Cool &  M5.5 & M16b  & -0.97 (-1.07,-0.87) \\ 
J11071206-7632232 &          T24 &       &      &  M0.5 &    II &   Hot &    M0 & M16a  & -0.23 (-0.34,-0.12) \\ 
J11071330-7743498 &      CHXR22E &       &      &  M3.5 &  II/T &   Hot &    M4 &  M14  & -0.63 (-0.71,-0.55) \\ 
J11071860-7732516 &       ChaH$\alpha$9 &       &      &  M5.5 &    II &  Cool &  M5.5 & M16a  & -0.92 (-1.02,-0.82) \\ 
J11072074-7738073 &          T26 & 4.570 &  D13 &    G2 &    II &   Hot &    K0 & M16b  &  0.29 ( 0.23, 0.56) \\ 
J11072825-7652118 &          T27 & 0.780 &  D13 &    M3 &    II &   Hot &    M3 & M16b  & -0.53 (-0.59,-0.47) \\ 
J11074245-7733593 &       ChaH$\alpha$2 & 0.167 & La08 & M5.25 &    II &  Cool &  M5.5 & M16b  & -0.88 (-0.98,-0.77) \\ 
J11074366-7739411 &          T28 & 28.87 & KH07 &    M0 &    II &   Hot &    M1 & M16b  & -0.31 (-0.43,-0.19) \\ 
J11074656-7615174 &    CHSM10862 &       &      & M5.75 &    II &  Cool &  M6.5 & M16b  & -1.15 (-1.25,-1.05) \\ 
J11075730-7717262 &      CHXR30B &       &      & M1.25 &    II &   Hot & M1.25 &  L07  & -0.31 (-0.43,-0.18) \\ 
J11075792-7738449 &         Sz22 & 0.500 &  G97 &    K6 &    FS &   Hot &    K5 & M16a  & -0.01 (-0.06, 0.23) \\ 
J11075809-7742413\tablenotemark{\ast} &          T30 &       &      &  M2.5 &    II &   Hot &    M3 & M16b  & -0.51 (-0.58,-0.44) \\ 
J11080002-7717304 &      CHXR30A & 0.460 & La08 &    K8 &    II &   Hot &    K7 & M16b  & -0.18 (-0.28,-0.07) \\ 
J11080148-7742288 &     VWCha,T31 & 0.660 &  D13 &    K8 &    II &   Hot &    K7 & M16a  & -0.20 (-0.30,-0.11) \\ 
J11080297-7738425 &     ESOH$\alpha$562 & 0.280 &  D13 & M1.25 &    FS &   Hot &    M1 & M16a  & -0.20 (-0.29, 0.04) \\ 
J11081509-7733531 &         T33A & 2.400 &  D13 &    G7 &    FS &   Hot &    K0 & M16a  &  0.12 ( 0.08, 0.16) \\ 
J11081850-7730408\tablenotemark{\ast} &       ISO138 &       &      &  M6.5 &    II &  Cool &  M6.5 & M16b  & -1.14 (-1.24,-1.03) \\ 
J11082238-7730277 &       ISO143 & 18.16 & KH07 &    M5 &    II &  Cool &  M5.5 & M16a  & -0.90 (-0.99,-0.79) \\ 
J11082570-7716396 &              &       &      &    M8 &    II &  Cool &    M8 &  L07  & -1.51\tablenotemark{\dagger} \\ 
J11082650-7715550 &       ISO147 &       &      & M5.75 &    II &  Cool &  M5.5 & M16b  & -0.96 (-1.06,-0.86) \\ 
J11083905-7716042 &         Sz27 &       &      &    K8 &  II/T &   Hot &    K7 &  M14  & -0.08 (-0.15, 0.16) \\ 
J11083952-7734166 &       ChaH$\alpha$6 &       &      & M5.75 &    II &  Cool &  M6.5 & M16a  & -0.99 (-1.06,-0.92) \\ 
J11085090-7625135 &          T37 &       &      & M5.25 &    II &  Cool &  M5.5 & M16b  & -0.90 (-0.99,-0.79) \\ 
J11085367-7521359 &              &       &      &  M1.5 &    II &   Hot &    M1 & M16b  & -0.28 (-0.39,-0.16) \\ 
J11085464-7702129 &          T38 &       &      &  M0.5 &    II &   Hot &  M0.5 & M16a  & -0.18 (-0.26, 0.06) \\ 
J11085497-7632410 &       ISO165 &       &      &  M5.5 &    II &  Cool &  M5.5 & M16b  & -0.91 (-1.00,-0.81) \\ 
J11091812-7630292 &       CHXR79 & 0.880 &  D13 & M1.25 &    II &   Hot &    M0 & M16b  & -0.18 (-0.28,-0.07) \\ 
J11092266-7634320 &         C1-6 &       &      & M1.25 &    II &   Hot &    M1 & M16b  & -0.25 (-0.33,-0.01) \\ 
J11092379-7623207 &          T40 &       &      &    K6 &    II &   Hot &  M0.5 & M16a  & -0.29 (-0.40,-0.18) \\ 
J11094260-7725578 &         C7-1 &       &      &    M5 &    II &  Cool &    M5 &  L07  & -0.77\tablenotemark{\dagger} \\ 
J11094621-7634463 &        Hn10e & 19.17 & KH07 & M3.25 &    II &   Hot &    M3 & M16b  & -0.47 (-0.54,-0.39) \\ 
J11094742-7726290 &   B43,ISO207 &       &      & M3.25 &    II &   Hot &    M1 & M16b  & -0.22 (-0.30, 0.02) \\ 
J11095215-7639128 &       ISO217 &       &      & M6.25 &    II &  Cool & M6.25 &  L07  & -1.20 (-1.76,-1.08) \\ 
J11095336-7728365 &       ISO220 &       &      & M5.75 &    II &  Cool &  M5.5 & M16b  & -0.96 (-1.06,-0.86) \\ 
J11095340-7634255 &          T42 &       &      &    K5 &    II &   Hot &    K7 & M16b  & -0.12 (-0.21,-0.03) \\ 
J11095407-7629253 &          T43 & 0.780 &  D13 &    M2 &    II &   Hot &    M1 & M16b  & -0.21 (-0.30, 0.03) \\ 
J11095873-7737088 &        WXCha & 0.740 &  D13 & M1.25 &    II &   Hot &  M0.5 & M16b  & -0.29 (-0.39,-0.19) \\ 
J11100010-7634578 &        WWCha & 0.006 &  A15 &    K5 &    II &   Hot &    K0 & M16a  &  0.21 ( 0.17, 0.27) \\ 
J11100369-7633291 &         Hn11 &       &      &    K8 &    II &   Hot &    M0 & M16b  & -0.14 (-0.23, 0.12) \\ 
J11100469-7635452 &    FNCha &       &      &    M1 &    II &   Hot &    K7 & M16a  & -0.08 (-0.15, 0.16) \\ 
J11100704-7629376 &          T46 & 0.120 &  N12 &    M0 &    II &   Hot &    K7 & M16b  & -0.14 (-0.24,-0.05) \\ 
J11100785-7727480 &       ISO235 &       &      &  M5.5 &    II &  Cool &  M5.5 & M16b  & -0.89 (-0.99,-0.79) \\ 
J11101141-7635292 &       ISO237 & 28.32 & KH07 &  K5.5 &    II &   Hot &    K5 & M16a  &  0.00 (-0.06, 0.26) \\ 
J11103801-7732399 &       CHXR47 & 0.170 &  D13 &    K3 &    II &   Hot &    K4 & M16b  &  0.05 (-0.04, 0.14) \\ 
J11104141-7720480 &       ISO252 &       &      &    M6 &    II &  Cool &  M5.5 & M16b  & -0.96 (-1.06,-0.86) \\ 
J11104959-7717517 &          T47 & 12.09 & KH07 &    M2 &    II &   Hot &    M2 & M16b  & -0.38 (-0.51,-0.25) \\ 
J11105333-7634319 &          T48 &       &      & M3.75 &    II &   Hot &    M3 & M16b  & -0.51 (-0.58,-0.44) \\ 
J11105359-7725004 &       ISO256 &       &      &  M4.5 &    II &  Cool &    M5 & M16b  & -0.81 (-0.90,-0.72) \\ 
J11105597-7645325 &         Hn13 & 0.130 & La08 & M5.75 &    II &  Cool &  M6.5 & M16b  & -0.98\tablenotemark{\dagger} \\ 
J11111083-7641574 &     ESOH$\alpha$569 &       &      &  M2.5 &    II &   Hot &    M1 & M16b  & -0.31\tablenotemark{\dagger} \\ 
J11113965-7620152 &          T49 & 24.38 & KH07 &    M2 &    II &   Hot &  M3.5 & M16a  & -0.59 (-0.65,-0.53) \\ 
J11114632-7620092 &      CHXN18N &       &      &    K6 &    II &   Hot &    K2 & M16a  &  0.09 ( 0.05, 0.13) \\ 
J11120351-7726009 &       ISO282 &       &      & M4.75 &    II &  Cool &  M5.5 & M16b  & -0.89 (-0.98,-0.81) \\ 
J11120984-7634366 &          T50 &       &      &    M5 &    II &  Cool &    M5 & M16b  & -0.78 (-0.84,-0.72) \\ 
J11122441-7637064 &          T51 & 1.970 &  D13 &  K3.5 &    II &   Hot &    K2 & M16a  &  0.04 ( 0.00, 0.10) \\ 
J11122772-7644223 &          T52 & 11.18 & KH07 &    G9 &    II &   Hot &    K0 & M16a  &  0.20 ( 0.15, 0.25) \\ 
J11123092-7644241\tablenotemark{\ast} &          T53 &       &      &    M1 &    II &   Hot &  M0.5 & M16b  & -0.17 (-0.26, 0.09) \\ 
J11124268-7722230 &         T54A & 0.240 &  D13 &    G8 &  II/T &   Hot &    K0 & M16a  &  0.20 ( 0.16, 0.25) \\ 
J11124861-7647066 &         Hn17 &       &      &    M4 &    II &  Cool &  M4.5 & M16a  & -0.69 (-0.77,-0.60) \\ 
J11132446-7629227\tablenotemark{\ast} &         Hn18 &       &      &  M3.5 &    II &   Hot &    M4 & M16a  & -0.62 (-0.69,-0.54) \\ 
J11142454-7733062 &        Hn21W & 5.480 &  D13 &    M4 &    II &  Cool &  M4.5 & M16a  & -0.71 (-0.79,-0.63) \\ 
J11160287-7624533 & ESOH$\alpha$574 &       &      &    K8 &    II &   Hot &    K8 &  L07  & -0.19\tablenotemark{\dagger} \\ 
J11173700-7704381 &         Sz45 &       &      &  M0.5 &  II/T &   Hot &  M0.5 &  M14  & -0.28 (-0.39,-0.17) \\ 
J11175211-7629392 &              &       &      &  M4.5 &    II &  Cool &  M4.5 &  L07  & -0.69 (-0.77,-0.61) \\ 
J11183572-7935548 &              &       &      & M4.75 &  II/T &  Cool &    M5 & M16b  & -0.77\tablenotemark{\dagger} \\ 
J11241186-7630425 &              &       &      &    M5 &  II/T &  Cool &  M5.5 & M16b  & -0.90 (-0.99,-0.79) \\ 
J11432669-7804454 &              &       &      &    M5 &    II &  Cool &  M5.5 & M16b  & -0.86 (-0.93,-0.79) \\ 
\enddata
\tablenotetext{\dagger}{For these stars we fixed the isochrone, hence there are no uncertainties associated with the estimated stellar mass, see Section~\ref{sect:stellar_masses}.}
\tablenotetext{\ast}{T30 is the secondary of T31 at a separation of 16\farcs52.
ISO~138 is the secondary of ISO~143 at 18\farcs16. T53 is the secondary of T52 at 11\farcs18.
Hn18 is the secondary of CHXR60 (not included in our ALMA survey) at a separation of 28\farcs28.}
\tablerefs{(A15) Anthonioz et al. 2015; (D13) Daemgen et al. 2013; 
(KH07) Kraus \& Hillenbrand 2007; (La08) Lafreniere et al. 2008;
(L07) Luhman 2007;  (M14) Manara et al. 2014; (M16a);
Manara et al. 2016a; (M16b) Manara et al. 2016b; (N12)  Nguyen et al. 2012; (S13) Schmidt et al. 2013
}
\end{deluxetable}

\begin{deluxetable}{lcccccc}
\tabletypesize{\scriptsize}
\tablecaption{ALMA Observations\label{tab:ALMAObservations}}
\tablewidth{0pt}
\tablehead{
\colhead{UTC Date} & \colhead{Number} & \colhead{Baseline Range} & \colhead{pwv} & \multicolumn{3}{c}{Calibrators}\\ 
\colhead{ } & \colhead{Antennas} & \colhead{(m)} & \colhead{(mm)} & \colhead{Flux} &  \colhead{Passband}  &  \colhead{Phase}  
}
\startdata
2014 May 1-3 & 37 & 17-558 & 0.6 & Pallas & J1427-4206 & J1058-8003 \\
2015 May 18-19 & 39 & 21-556 & 0.6 & Ganymede,J1107-448 & J0538-4405,J1337-1257 & J1058-8003 \\
\enddata
\end{deluxetable}

\begin{deluxetable}{lcccc}
\tabletypesize{\tiny}
\tablecaption{Measured Continuum Flux Densities\label{tab:SourceFluxes}}
\tablewidth{0pt}
\tablehead{
\colhead{2MASS} & \colhead{F$_\nu$} & \colhead{$\Delta \alpha$} &  \colhead{$\Delta \delta$} & \colhead{FWHM} \\ 
\colhead{ } & \colhead{(mJy)} & \colhead{(arcsec)} & \colhead{(arcsec)} & \colhead{(arcsec)} 
}
\startdata
J10533978-7712338 &    4.60$\pm$0.79 & -0.37$\pm$0.03 & -0.09$\pm$0.05 &  ... \\ 
J10555973-7724399 &   34.10$\pm$1.32 & -0.06$\pm$0.01 & -0.14$\pm$0.01 &  0.18$\times$0.11 \\ 
J10561638-7630530 &    3.99$\pm$0.16 & -0.37$\pm$0.01 & -0.08$\pm$0.01 &  ... \\ 
J10563044-7711393\tablenotemark{\dagger} &  117.58$\pm$1.10 &  ... &  ... &  ... \\ 
J10574219-7659356 &    9.12$\pm$0.83 & -0.27$\pm$0.02 &  0.04$\pm$0.03 &  ... \\ 
J10580597-7711501 &    2.68$\pm$0.16 & -0.38$\pm$0.01 & -0.01$\pm$0.02 &  ... \\ 
J10581677-7717170\tablenotemark{\dagger} &  310.18$\pm$1.00 &  ... &  ... &  ... \\ 
J10590108-7722407 &   65.34$\pm$1.70 & -0.40$\pm$0.01 & -0.15$\pm$0.01 &  0.40$\times$0.33 \\ 
J10590699-7701404 &  442.18$\pm$0.76 & -0.39$\pm$0.00 &  0.21$\pm$0.00 &  0.51$\times$0.44 \\ 
J11004022-7619280 &   69.75$\pm$0.17 & -0.28$\pm$0.00 & -0.01$\pm$0.00 &  0.38$\times$ 0.35 \\ 
J11022491-7733357 &  225.68$\pm$0.74 & -0.44$\pm$0.00 &  0.15$\pm$0.00 &  0.45$\times$0.45 \\ 
J11023265-7729129 &   -0.21$\pm$0.82 &  ... &  ... &  ... \\ 
J11025504-7721508 &    1.16$\pm$0.16 & -0.41$\pm$0.03 &  0.04$\pm$0.04 &  ... \\ 
J11040425-7639328 &    2.77$\pm$0.16 & -0.39$\pm$0.01 & -0.11$\pm$0.02 &  ... \\ 
J11040909-7627193 &  104.78$\pm$0.60 & -0.25$\pm$0.00 &  0.03$\pm$0.00 &  0.25$\times$0.25 \\ 
J11044258-7741571\tablenotemark{\ast} &    4.15$\pm$0.16 & -0.29$\pm$0.01 & -0.03$\pm$0.01 &  ... \\ 
J11045701-7715569 &    2.54$\pm$0.81 & -0.32$\pm$0.07 & -0.01$\pm$0.09 &  ... \\ 
J11062554-7633418 &   46.05$\pm$0.15 & -0.43$\pm$0.00 & -0.05$\pm$0.00 &  0.23$\times$0.15 \\ 
J11062942-7724586 &    0.25$\pm$0.16 &  ... &  ... &  ... \\ 
J11063276-7625210 &   -0.01$\pm$0.16 &  ... &  ... &  ... \\ 
J11063945-7736052 &    0.37$\pm$0.16 &  ... &  ... &  ... \\ 
J11064180-7635489 &    0.97$\pm$0.16 & -0.37$\pm$0.03 & -0.23$\pm$0.05 &  ... \\ 
J11064510-7727023 &    0.53$\pm$0.82 &  ... &  ... &  ... \\ 
J11065906-7718535 &   24.28$\pm$0.35 & -0.47$\pm$0.00 &  0.12$\pm$0.00 &  0.17$\times$0.15 \\ 
J11065939-7530559 &    3.11$\pm$0.16 & -0.30$\pm$0.01 & -0.04$\pm$0.01 &  ... \\ 
J11070925-7718471 &    0.06$\pm$0.82 &  ... &  ... &  ... \\ 
J11071181-7625501 &    0.03$\pm$0.16 &  ... &  ... &  ... \\ 
J11071206-7632232 &    4.23$\pm$0.81 & -0.42$\pm$0.04 & -0.03$\pm$0.06 &  ... \\ 
J11071330-7743498 &    0.42$\pm$0.81 &  ... &  ... &  ... \\ 
J11071860-7732516 &    0.93$\pm$0.16 & -0.46$\pm$0.03 &  0.12$\pm$0.05 &  ... \\ 
J11072074-7738073 &   26.36$\pm$1.46 & -0.31$\pm$0.01 &  0.02$\pm$0.01 &  0.29$\times$0.29 \\ 
J11072825-7652118 &    1.50$\pm$0.81 &  ... &  ... &  ... \\ 
J11074245-7733593 &    2.37$\pm$0.41 & -0.40$\pm$0.03 & -0.02$\pm$0.04 &  0.34$\times$0.30 \\ 
J11074366-7739411 &  107.27$\pm$0.56 & -0.28$\pm$0.00 &  0.02$\pm$0.00 &  0.26$\times$0.19 \\ 
J11074656-7615174 &    2.18$\pm$0.16 & -0.31$\pm$0.01 &  0.09$\pm$0.02 &  ... \\ 
J11075730-7717262 &    6.47$\pm$0.80 & -0.34$\pm$0.03 &  0.02$\pm$0.04 &  ... \\ 
J11075792-7738449 &   19.85$\pm$1.48 & -0.18$\pm$0.01 &  0.00$\pm$0.02 &  0.33$\times$0.33 \\ 
J11075809-7742413 &    6.45$\pm$0.79 & -0.23$\pm$0.03 & -0.11$\pm$0.04 &  ... \\ 
J11080002-7717304 &   -0.69$\pm$0.80 &  ... &  ... &  ... \\ 
J11080148-7742288 &   44.37$\pm$0.82 & -0.20$\pm$0.00 &  0.31$\pm$0.01 &  ... \\ 
J11080297-7738425 &  102.24$\pm$0.58 & -0.30$\pm$0.00 & -0.07$\pm$0.00 &  0.28$\times$0.28 \\ 
J11081509-7733531 &  209.29$\pm$0.43 &  1.00$\pm$0.00 & -0.26$\pm$0.00 &  0.46$\times$0.46 \\ 
J11081850-7730408 &    0.26$\pm$0.16 &  ... &  ... &  ... \\ 
J11082238-7730277\tablenotemark{\ast} &    0.23$\pm$0.16 &  ... &  ... &  ... \\ 
J11082570-7716396 &    0.23$\pm$0.15 &  ... &  ... &  ... \\ 
J11082650-7715550 &   -0.24$\pm$0.16 &  ... &  ... &  ... \\ 
J11083905-7716042 &   14.11$\pm$0.79 & -0.36$\pm$0.01 &  0.04$\pm$0.02 &  ... \\ 
J11083952-7734166 &    0.02$\pm$0.16 &  ... &  ... &  ... \\ 
J11085090-7625135 &   -0.04$\pm$0.16 &  ... &  ... &  ... \\ 
J11085367-7521359 &   24.60$\pm$1.37 & -0.21$\pm$0.01 &  0.01$\pm$0.01 &  0.30$\times$0.22 \\ 
J11085464-7702129 &    3.90$\pm$0.79 & -0.38$\pm$0.04 & -0.05$\pm$0.06 &  ... \\ 
J11085497-7632410 &    0.46$\pm$0.16 &  ... &  ... &  ... \\ 
J11091812-7630292 &    1.30$\pm$0.79 &  ... &  ... &  ... \\ 
J11092266-7634320 &    3.85$\pm$0.78 & -0.24$\pm$0.04 & -0.26$\pm$0.06 &  ... \\ 
J11092379-7623207 &  123.11$\pm$0.57 & -0.30$\pm$0.00 & -0.07$\pm$0.00 &  0.26$\times$ 0.23 \\ 
J11094260-7725578 &    0.37$\pm$0.16 &  ... &  ... &  ... \\ 
J11094621-7634463 &    4.73$\pm$0.79 & -0.37$\pm$0.03 & -0.08$\pm$0.05 &  ... \\ 
J11094742-7726290 &  147.85$\pm$0.86 & -0.42$\pm$0.00 & -0.08$\pm$0.00 &  0.75$\times$0.45 \\ 
J11095215-7639128 &    0.37$\pm$0.16 &  ... &  ... &  ... \\ 
J11095336-7728365 &    0.29$\pm$0.16 &  ... &  ... &  ... \\ 
J11095340-7634255 &   76.10$\pm$1.83 & -0.34$\pm$0.01 &  0.06$\pm$0.01 &  0.56$\times$0.38 \\ 
J11095407-7629253 &   30.49$\pm$1.24 & -0.36$\pm$0.01 & -0.17$\pm$0.01 &  0.13$\times$0.13 \\ 
J11095873-7737088 &   20.81$\pm$0.57 & -0.53$\pm$0.01 & -0.10$\pm$0.01 &  ... \\ 
J11100010-7634578 & 1363.47$\pm$0.82 & -0.40$\pm$0.00 & -0.03$\pm$0.00 &  0.56$\times$0.44 \\ 
J11100369-7633291 &    9.83$\pm$0.79 & -0.32$\pm$0.02 & -0.02$\pm$0.02 &  ... \\ 
J11100469-7635452 &    7.73$\pm$0.78 & -0.36$\pm$0.02 &  0.04$\pm$0.03 &  ... \\ 
J11100704-7629376 &    7.17$\pm$0.78 & -0.35$\pm$0.02 &  0.09$\pm$0.03 &  ... \\ 
J11100785-7727480 &    0.49$\pm$0.16 &  ... &  ... &  ... \\ 
J11101141-7635292 &   73.82$\pm$1.40 & -0.41$\pm$0.00 &  0.07$\pm$0.00 &  0.19$\times$0.16 \\ 
J11103801-7732399 &    5.37$\pm$0.78 & -0.26$\pm$0.03 &  0.09$\pm$0.04 &  ... \\ 
J11104141-7720480 &   -0.00$\pm$0.16 &  ... &  ... &  ... \\ 
J11104959-7717517 &   58.37$\pm$1.45 & -0.41$\pm$0.00 &  0.11$\pm$0.01 &  0.34$\times$0.21 \\ 
J11105333-7634319 &   31.02$\pm$1.29 & -0.33$\pm$0.01 &  0.01$\pm$0.01 &  0.22$\times$0.18 \\ 
J11105359-7725004 &    7.88$\pm$0.34 & -0.38$\pm$0.01 & -0.07$\pm$0.01 &  0.18$\times$0.13 \\ 
J11105597-7645325 &    2.23$\pm$0.22 & -0.41$\pm$0.02 & -0.08$\pm$0.03 &  ... \\ 
J11111083-7641574 &   54.27$\pm$1.75 & -0.32$\pm$0.01 &  0.34$\pm$0.01 &  0.74$\times$0.16 \\ 
J11113965-7620152 &   21.48$\pm$0.80 & -0.25$\pm$0.01 &  0.38$\pm$0.01 &  ... \\ 
J11114632-7620092 &   35.20$\pm$1.26 & -0.25$\pm$0.01 &  0.43$\pm$0.01 &  0.19$\times$0.11 \\ 
J11120351-7726009 &    2.95$\pm$0.16 & -0.42$\pm$0.01 &  0.07$\pm$0.02 &  ... \\ 
J11120984-7634366 &    4.44$\pm$0.22 & -0.37$\pm$0.01 &  0.07$\pm$0.01 &  ... \\ 
J11122441-7637064 &    0.19$\pm$0.78 &  ... &  ... &  ... \\ 
J11122772-7644223 &   59.05$\pm$1.29 & -0.38$\pm$0.00 &  0.12$\pm$0.00 &  0.12$\times$0.10 \\ 
J11123092-7644241\tablenotemark{\ast} &   12.18$\pm$0.83 & -0.32$\pm$0.01 &  0.25$\pm$0.02 &  ... \\ 
J11124268-7722230 &   -0.02$\pm$0.79 &  ... &  ... &  ... \\ 
J11124861-7647066 &   -0.10$\pm$0.16 &  ... &  ... &  ... \\ 
J11132446-7629227 &    8.07$\pm$0.79 & -0.35$\pm$0.02 & -0.05$\pm$0.03 &  ... \\ 
J11142454-7733062 &    7.43$\pm$0.34 & -0.40$\pm$0.01 & -0.19$\pm$0.01 &  0.18$\times$0.16 \\ 
J11160287-7624533 &   12.83$\pm$1.68 & -0.34$\pm$0.03 &  0.06$\pm$0.04 &  0.43$\times$0.43 \\ 
J11173700-7704381 &   28.26$\pm$1.29 & -0.36$\pm$0.01 & -0.03$\pm$0.01 &  0.20$\times$0.20 \\ 
J11175211-7629392 &   -0.31$\pm$0.16 &  ... &  ... &  ... \\ 
J11183572-7935548 &   14.52$\pm$0.35 &  0.18$\pm$0.00 & -0.32$\pm$0.01 &  0.22$\times$0.22 \\ 
J11241186-7630425 &    1.47$\pm$0.16 & -0.35$\pm$0.02 & -0.12$\pm$0.03 &  ... \\ 
J11432669-7804454 &    1.36$\pm$0.50 & -0.51$\pm$0.10 & -0.24$\pm$0.11 &  0.53$\times$0.50 \\ 
\enddata
\tablecomments{Sources with a FWHM reported in the last column of the table are those that were fitted with an elliptical gaussian.
Undetected sources have ellipses in all columns following the flux density column. For these sources flux densities are measured
assuming a point source model and fixed $\Delta \alpha$ and $\Delta \delta$ to the median values of the detected sources.}
\tablenotetext{\dagger}{Sources with rings. Integrated flux density measured on image within the 3$\sigma$ contour.}
\tablenotetext{\ast}{Sources that have additional mm detections in their exposures: J11044258-7741571 (ISO~52) at $\sim$6\arcsec{}, coordinates (11:04:40.59;-77:41:56.9); J11082238-7730277 (ISO~143) at $\sim$10\arcsec{}, coordinates (11:08:21.11;-77:30:18.9); and J11123092-7644241 (T53) at $\sim 11$\arcsec{}, coordinates (11:12:27.7;-76:44:22.3). In the first two cases there is no object in the SIMBAD Astronomical Database associated with the mm emission. In the case of T53 we detect the disk from the companion T52. Fluxes from these additional detections are not reported in the table. }

\end{deluxetable}

\begin{deluxetable}{lc cc cc c}
\tabletypesize{\tiny}
\tablecaption{$M_{\rm dust}-M_*$ relations.\label{tab:relations}}
\tablewidth{0pt}
\tablehead{
\colhead{Region} & \colhead{Age (Myr)} & \colhead{$\alpha_{T_{20}}$} & \colhead{$\beta_{T_{20}}$}  & \colhead{$\alpha$} & \colhead{$\beta$} & \colhead{Dispersion}
}
\startdata
Taurus &  1-2 & 1.6(0.2) & 1.2(0.1) & 1.1(0.2) & 1.0(0.1) & 0.7(0.1) \\
Lupus\tablenotemark{\ast} & 1-3 & 1.8(0.3) & 1.6(0.2)& 1.1(0.3) & 1.4(0.2) & 0.8(0.1)\\
Cha~I & 2-3 & 1.9(0.2)& 1.1(0.1) & 1.3(0.2) & 1.1(0.1) & 0.8(0.1) \\
Upper~Sco & 10 & 2.7(0.4)& 0.9(0.2) &  1.9(0.4) & 0.8(0.2) & 0.7(0.1) \\
\enddata
\tablecomments{The listed $\alpha$ and $\beta$ values (uncertainties in parenthesis) are the slope and intercept of the following linear relation: log($M_{\rm dust}/M_\Earth$)=$\alpha \times$log($M_*/M_\sun$)+$\beta$. The first two entries are obtained assuming a fixed dust temperature of 20\,K while the other entries assume a dust disk temperature scaling with stellar luminosity (see text for more details).}
\tablenotetext{\ast}{There are twenty sources in Lupus that do not have stellar masses \citep{ansdell16,alcala16}. While \citet{ansdell16} have assigned masses in a MC fashion following the distribution of the Lupus I-IV YSOs, we do not include these sources in our fits. The slope and dispersion reported in \citet{ansdell16} are the same as those reported here. }
\end{deluxetable}

\begin{deluxetable}{l cc c cc}
\tabletypesize{\tiny}
\tablecaption{Summary of methods for the $F_{\rm mm}-M_*$ relation in \ChaI .\label{tab:tests}}
\tablewidth{0pt}
\tablehead{
\colhead{Routine} & \colhead{Method} & \colhead{Censored} & \colhead{Slope}  & \colhead{Intercept} 
}
\startdata
fitexy (IDL) &  Nukers                 & n   & 2.43(0.08) & 2.16(0.04) \\
mpfitexy (IDL) &  Nukers (with scatter) &  n &    1.5(0.2) & 1.8(0.1) \\
censReg (R) & Maximum Likelihood & y & 1.8(0.2) & 1.6(0.1) \\
cenken (R) & Akritas-Thiel-Sen & y & 1.9 & 1.7 \\
linmix\_err (IDL) & Bayesian  & y &  1.9(0.2) & 1.6(0.1) \\
\enddata
\tablecomments{Uncertainties in the slope and intercept are reported in parenthesis.}
\end{deluxetable}

\clearpage


\end{document}